\documentclass{jaa}
\usepackage[authoryear]{natbib}
\usepackage{microtype}
\usepackage{booktabs}
\usepackage{xcolor}
\usepackage{newtxtext,newtxmath}
\usepackage[T1]{fontenc}
\usepackage{microtype}
\usepackage{amsmath}	
\usepackage{amstext,amscd,bm}
\usepackage{array}
\usepackage{multirow}

\renewcommand{\thesection}{\arabic{section}}

\usepackage{graphicx}
\usepackage[colorlinks=true,citecolor=blue,linkcolor=blue,urlcolor=blue]{hyperref}

\begin{document}
	
	\title
	{CCD $UBV(RI)_{KC}$ photometry of the open cluster NGC~6793 and its dynamical evolution}


	\author{Zahra Al\textsuperscript{1,*}, 
		Y\"uksel Karata\c{s}\textsuperscript{2}, 
		Ra\'ul Michel\textsuperscript{3}, 
		Charles Bonatto\textsuperscript{4}, 
		Orhan G\"une\c{s}\textsuperscript{5}, 
		{\.I}nci Akkaya Oralhan\textsuperscript{6}, 
		Ey\"up Kaan \"Ulgen\textsuperscript{7}
	}
	
	\affilOne{\textsuperscript{1}Institute of Graduate Studies In Sciences, {\.I}stanbul University, 34116, \"Universite-Istanbul, T\"urkiye.}
	\affilTwo{\textsuperscript{2}Department of Astronomy and Space Sciences, Science Faculty, {\.I}stanbul University, 34116, \"Universite-Istanbul, T\"urkiye.}
	\affilThree{\textsuperscript{3}Observatorio Astron\'omico Nacional, Universidad Nacional Aut\'onoma de M\'exico, Apartado Postal 877, C.P. 22800, Ensenada, B.C., M\'exico.}
	\affilFour{\textsuperscript{4}Universidade Federal do Rio Grande do Sul, Departamento de Astronomia, CP\,15051, RS, Porto Alegre 91501-970, Brazil.}
	\affilFive{\textsuperscript{5}Department of Science History, Faculty of Arts and Sciences, Medeniyet University, G\"oztepe, 34720, Istanbul, T\"urkiye.}
	\affilSix{\textsuperscript{6}Department of Astronomy and Space Sciences, Faculty of Arts and Sciences, Erciyes University, Talas Yolu, 38039, Kayseri, T\"urkiye.}
	\affilSeven{\textsuperscript{7}Huawei T\"urkiye Research and Development Center, 34768, {\.I}stanbul, T\"urkiye.}
	
	\twocolumn[{
		\maketitle
		\corres{zahraa96al@gmail.com}

		\msinfo{1 January 2026}{1 January 2026}
		
		
		\begin{abstract}
				We present new astrophysical parameters for the open cluster NGC~6793 based on new CCD $UBV(RI)_{KC}$ photometry. We derived a reddening of $E(B-V) = 0.24 \pm 0.02$~mag and a heavy element abundance of $Z = 0.024$ ($[Fe/H] = +0.20$~dex). Padova isochrone fitting to the $V \times (B-V)$ colour-magnitude diagram yields an intermediate age of $525 \pm 51$~Myr and a distance modulus of $\mu = 8.80 \pm 0.05$~mag, corresponding to a distance of $d = 575 \pm 58$~pc from the Sun. The core radius of NGC~6793 appears to be shrinking due to advanced dynamical evolution ($\log\tau_{2} = 1.13$), driven by mass segregation and the evaporation of low-mass stars from the central region. The ratios of core to half-mass radius ($R_{c}/R_{h}$) and half-mass to Jacobi radius ($R_{h}/R_{J}$) indicate that the cluster's evolution is governed by the combined effects of internal two-body relaxation, mass segregation, and external tidal perturbations. The ratio $R_{t}/R_{J} = 0.99$ suggests that the cluster is currently in a tidally filling state. The parameter pairs ($t_{diss}/t_{rlx_{1}} = 40$, $\log R_{J}/R_{c} = 0.72$) and ($R_{h}/R_{J} = 0.38$, $\log\rho_{amb} = -0.88$) place NGC~6793 among the relatively compact clusters within $R_{GC} < 7.9$~kpc. This implies a compact internal structure that is stable against the combined effects of two-body encounters and tidal heating. Given its current state, NGC~6793 will likely dissolve and disperse before entering the final contraction phase ($R_{4}$ regime).
			
		\end{abstract}

		\keywords{(Galaxy:) open clusters and associations:general - Galaxy: abundances - Galaxy: evolution.}

	}]
	
	
	\doinum{12.3456/s78910-011-012-3}
	\artcitid{\#\#\#\#}
	\volnum{000}
	\year{0000}
	\pgrange{1--}
	\setcounter{page}{1}
	\lp{1}

	\section{Introduction}\label{intro}

	NGC~6793 open cluster (OC) resides in the first Galactic quadrant and in the constellation of Vulpecula. NGC~6793 is poorly studied in UBVRI filters but it was extensively studied with Gaia data (See Tables 3 and 8).  Thus, we present its new astrophysical parameters, reddening E(B-V), distance modulus ($\mu$), distance (d) and Age from new CCD~$UBV(RI)_{KC}$ photometry within the Sierra San Pedro M\'artir National Astronomical Observatory  Open Cluster Survey (SPMO, hereafter). In conjunction with this, SPMO presents a rare quality and reliable observations of the Galactic open clusters \citep{suc07, tap10}.  The photometric analyses towards the astrophysical parameters have been made with the help of the $(U-B)$, $(B-V)$ colour-colour (CC) and colour-magnitude diagrams (CMD) which are built with new CCD~$UBV(RI)_{KC}$ photometry. Its metal abundance for the isochrone selection has been obtained  from the photomtric metallicity calibration, $[Fe/H]-\delta_{0.6}$ of  \cite{karatas2006}.

	Gaia DR3 astrometric/photometric and radial velocity data have also been utilised for the separation of the cluster members, obtaining of the structural/dynamical parameters, and thus estimating kinematics, respectively. The dynamical parameters for NGC~6793 were derived by \cite{hunt2023} (hereafter H23), \citet{alm2023,alm2025} and \cite{Angelo2025} (hereafter A25) from Gaia DR3 data. We interprete its dynamical evolution in the light of new astrophysical parameters (distance and age) from new CCD~$UBV(RI)_{KC}$. The dynamical evolution of a individual OC like NGC~6793 can properly be interpreted from the trends between the dynamical indicators derived from larger sample OCs obtained in the Gaia epoch (A25). 
	
	The definition of dynamic parameters of the OCs is excellently presented as "terminology" in Appendix A of the paper of \cite{por2001}. \cite{kar2023}, \cite{Angelo2020,Angelo2021, Angelo2023}, A25 and \cite{lior2022} also provide a good summary of how to interpret the dynamic evolution indicators of the OCs in terms of their location in the Galaxy and time scales. Besides this, the information of dynamical evolution of the OCs in the light of internal and external processes can be found in the following  papers,  \cite{schil06}, \cite{Lamers2006,Gieles2007}, \cite{camargo2009,camargo2010}, \cite{Bon2010}, \cite{Gunes2017} hereafter G17), \cite{pia17a,pia17b,pia09}, \cite{Angelo2020, Angelo2021, Angelo2023}, A25, \cite{Tarr2022}, and \cite{akk24}.
	
	\cite{heg2003} have described the internal dynamics  evolution of a star cluster in the $R_{c}/R_{h}$ versus $R_{h}/R_{t}$ plane. As a star cluster expands to the point of being tidally filling, it experiences violent relaxation in its core region followed by two-body relaxation, mass segregation, and finally core-collapse. These processes are summarized in their Fig.2a. The most massive stars in an open cluster transfer/impart kinetic energy to the lower-mass stars. As a result, the centrally concentrated massive stars should have significantly lower velocities than fast-moving low-mass objects on the periphery of the cluster. This is known as energy equipartition.
	
	The organization of this paper is as follows: Section~\ref{sec-data} describes the data selection from Gaia DR3, observations at SPMO, data reduction, and the CCD $UBV(RI)_{KC}$ photometric analysis of NGC 6793. Structural parameters and the cluster membership technique are presented in Section~\ref{sec:structural_membership}. In Section~\ref{sec-astro}, we derive the reddening, metal abundance, distance modulus ($\mu$), and age using both CCD $UBV(RI)_{KC}$ and Gaia DR3 photometry. Kinematics, orbital parameters, mass function slope, dynamical parameters, time scales, and radial migration are detailed in Sections~\ref{sec-kinematics} through \ref{sec-dynamics}. Finally, a review of our findings in the context of existing literature is provided in the Discussion and Conclusion (Section~\ref{sec-discussion}).

	\begin{figure}[!t]
		\centering
		\includegraphics[width=1.0\columnwidth]{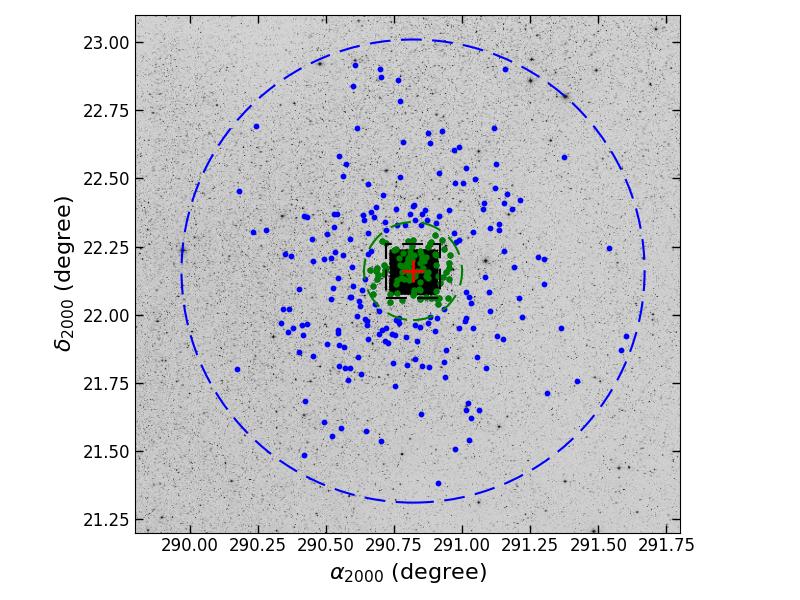}
		\caption{The star chart of NGC~6793 from the webpage of The STScI Digitized Sky Survey. 
			Black square shows the field of view of the SPM detector, $10^{\prime} (N-S)\times10^{\prime}(E-W)$ arcmin$^2$, 
			which corresponds to $9.543^{\prime} (N-S)\times8.813^{\prime}$ arcmin$^2$ as the lengths of the field in arcminutes. Black area and red plus denote 1283 cluster member and field stars with CCD UBVRI and the central equatorial coordinates (Table~\ref{tbl-1-summary}), respectively. 
			Green/blue circles and green/blue dots represent the core/limit radii (Cols.~3-4 of Table~3) and core/tidal cluster members, respectively. The analysis was performed using data within a $60^{\prime}$ circular area (the blue dashed line) centered on the cluster coordinates.}
		\label{fig-chart}
	\end{figure}

	\section{Observations and Data}
	\label{sec-data}
	
	\subsection{Gaia DR3}\label{sec-gaia-data}
	
	Gaia DR3 \citep{val2023} provides unprecedented astrometric and photometric data for nearly 1.3 billion sources, including celestial coordinates $(\alpha, \delta)$, trigonometric parallax, proper motions in right ascension and declination $(\mu_{\alpha}^*, \mu_{\delta})$, and broadband $G$, $G_{BP}$ and $G_{RP}$ photometry, with a magnitude range of $3 \lesssim G \lesssim 21$ \citep{val2023}. For this study, we retrieved astrometric and photometric parameters, along with radial velocities (where available), from Gaia ESA Archive\footnote{\url{https://gea.esac.esa.int/archive}} using \texttt{gaiadr3.gaia\_source} table. The table \texttt{gaiadr3.astrophysical\_parameters} was used to extract stellar atmospheric parameters—effective temperature ($T_{eff}$), metallicity ([M/H] and [Fe/M]), and surface gravity ($\log g$). These include estimates from the \texttt{GSP-Spec} module (based on higher-resolution Radial Velocity Spectrometer (RVS) spectra) \citealt{cropper2018} and \texttt{ESP-HS} module (Fr´emat et al. 2023) which provides data for $T_{eff}$ and $log g$. We note that iron abundance from \texttt{GSP-Spec} is inferred via $[Fe/H] = [Fe/M] + [M/H]$. Adopting the cluster center coordinates listed in Table~\ref{tab:coords}, we selected all Gaia DR3 sources within a circular radius of $60'$, yielding an initial sample of $663\,316$ stars. Following the approach of \citet{lindegren21}, we applied the quality filter RUWE $<$ 1.4 (where RUWE stands for Renormalized Unit Weight Error), reducing the sample to $633\,937$ stars, whose trigonometric parallaxes were corrected for the Gaia DR3 parallax zero-point offsets \citep{lin2021}\footnote{\url{https://gitlab.com/icc-ub/public/gaiadr3-zeropoint}}. The subsequent analysis of structural parameters and membership determination is performed on the filtered sample of $633\,937$ stars.
	
	\subsection{CCD Photometry}
	
\begin{figure}[!t]
		\centering
		\includegraphics[width=0.75\columnwidth]{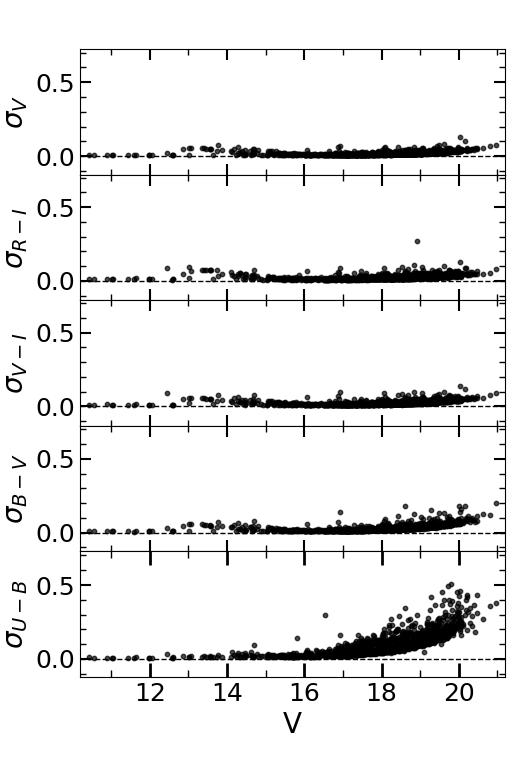}
		\caption{The distribution of the photometric errors of $V$, $(R-I)$, $(V-I)$, $(B-V)$ and $(U-B)$ against $V$ mag. All panels include 1283 cluster stars.}
		\label{fig-pherr}
	\end{figure}

	Observations of the open cluster NGC~6793, along with selected Landolt standard fields, were conducted at the Sierra San Pedro Mártir Observatory (SPMO) during clear photometric nights in May 2022. The 0.84-m (f/15) Ritchey–Chretien telescope was used, equipped with the Mexman filter wheel and a 2048 × 2048 pixel E2V CCD-231-42 detector (13.5~$\mu$m square pixels, gain of 2.38 e$^-$/ADU, readout noise 4.02 e$^-$). A 2 × 2 binning was applied, providing an unvignetted field of view of 10 × 10 arcmin$^2$. NGC~6793 was imaged through Johnson’s $UBV$ and Kron–Cousins $RI$ filters using both short and long exposures to properly capture bright and faint stars in the cluster region. A total of 1283 stars were detected in the cluster field. The corresponding star chart is shown in Fig.~\ref{fig-chart}. Standard fields \citep{lan09} were observed at zenith angles of approximately 60° and near the meridian to determine atmospheric extinction coefficients accurately. A log of observations, including the cluster’s central equatorial and Galactic coordinates, airmass range, and exposure times in each filter, is provided in Table1. Flat-field frames were obtained at the beginning and end of each night, while bias frames were recorded between cluster exposures. Data reduction was performed using the IRAF/DAOPHOT package {IRAF is distributed by the National Optical Observatories, operated by the Association of Universities for Research in Astronomy, Inc., under cooperative agreement with the National Science Foundation.} package \citep{stet87} by Raul Michel.
	
	\begin{equation}
		M_{\lambda} = m_{\lambda} - [k_{1\lambda} -k_{2\lambda}C)] X + \eta_{\lambda} C + \zeta_{\lambda}
	\end{equation} 
	
	where $m_{\lambda}$, $k_{1\lambda}$, $k_{2\lambda}$, $C$, and $X$  are the observed instrumental magnitude, primary/secondary extinction coefficients, colour index and air mass, respectively. $M_{\lambda}$, $\eta_{\lambda}$, $\zeta_{\lambda}$ are standard magnitude, transformation coefficient and photometric zero point, respectively. More details on data reduction together the extinction coefficients and zero points for $UBVRI$ filters can be found in the papers of \cite{akk10}, \cite{akk15} and \cite{akk19}. The photometric errors in $V$ and colours $(R-I)$, $(V-I)$, $(B-V)$, $(U-B)$ against  $V$ for  NGC~6793 are presented in Fig.~\ref{fig-pherr}, 1283 stars observed by SPMO. The mean errors for $V$-mag intervals are listed in Table~\ref{tab:photometric_errors}. The photometric errors in both magnitude and the colours for the range 10 $< V <$ 20 are less than $\sim0.10$ mag, except for $\sigma_{U-B}$. Larger errors of $\sigma_{U-B}$ as expected are seen at $V >$16.
	
	\renewcommand{\tabcolsep}{6mm}
	\renewcommand{\arraystretch}{1.1}
	\begin{table}[htb]
		\tabularfont
		\caption{Equatorial/Galactic coordinates and observation summary of NGC\,6793.}\label{tbl-1-summary}
		\begin{tabular}{lr}
			\topline
			$\alpha(2000)$\,(h\,m\,s)                              & 19 23 17.1~~$(290^{\circ}.82)$  \\
			$\delta(2000)\,(^{\circ}\,^{\prime}\,^{\prime\prime})$ & +22 09 28.5~~$(+22^{\circ}.16)$ \\
			$\ell\,(^{\circ})$                                     & 56.16   \\
			$b\,(^{\circ})$                                        & +3.32   \\
			Airmass                                                & 1.012 – 1.023 \\
			U (Exp.Time (s))                                       & 100, 1000 \\
			B (Exp.Time (s))                                       & 6, 60, 600 \\
			V (Exp.Time (s))                                       & 4, 40, 400 \\
			R (Exp.Time (s))                                       & 2, 20, 200 \\
			I (Exp.Time (s))                                       & 2, 20, 200 \\
			N                                                      & 1283 cluster stars  \\
			\midline
		\end{tabular}
		\tablenotes{Observation details including coordinates, filters, exposure times, and airmass.}
		\label{tab:coords}
	\end{table}

	\renewcommand{\tabcolsep}{3mm}
	\renewcommand{\arraystretch}{1.0}
	\begin{table}[htb]
		\tabularfont
		\caption{The mean photometric errors of $V$, $(R-I)$, $(V-I)$, $(B-V)$ and $(U-B)$ 
			for 1283 stars in the field of NGC~6793 in terms of $V$ mag.}\label{tab:photometric_errors}
		\begin{tabular}{lccccc}
			\topline
			$V$ & $\sigma_{V}$ & $\sigma_{R-I}$ & $\sigma_{V-I}$ & $\sigma_{B-V}$ & $\sigma_{U-B}$ \\
			\midline
			10--11 & 0.006 & 0.014 & 0.013 & 0.009 & 0.009 \\
			11--12 & 0.006 & 0.013 & 0.012 & 0.009 & 0.008 \\
			12--13 & 0.016 & 0.030 & 0.030 & 0.019 & 0.014 \\
			13--14 & 0.044 & 0.061 & 0.048 & 0.046 & 0.015 \\
			14--15 & 0.025 & 0.033 & 0.032 & 0.030 & 0.026 \\
			15--16 & 0.015 & 0.017 & 0.019 & 0.018 & 0.022 \\
			16--17 & 0.011 & 0.016 & 0.016 & 0.017 & 0.036 \\
			17--18 & 0.013 & 0.017 & 0.018 & 0.020 & 0.058 \\
			18--19 & 0.019 & 0.025 & 0.025 & 0.032 & 0.100 \\
			19--20 & 0.028 & 0.034 & 0.036 & 0.051 & 0.179 \\
			20--21 & 0.048 & 0.052 & 0.059 & 0.097 & 0.298 \\
			\hline
		\end{tabular}
	\end{table}

	\section{Structural parameters and Membership selection}\label{sec:structural_membership}
	
	\subsection{Structural Parameters}
\begin{figure}
		\centering
		\includegraphics[width=\columnwidth]{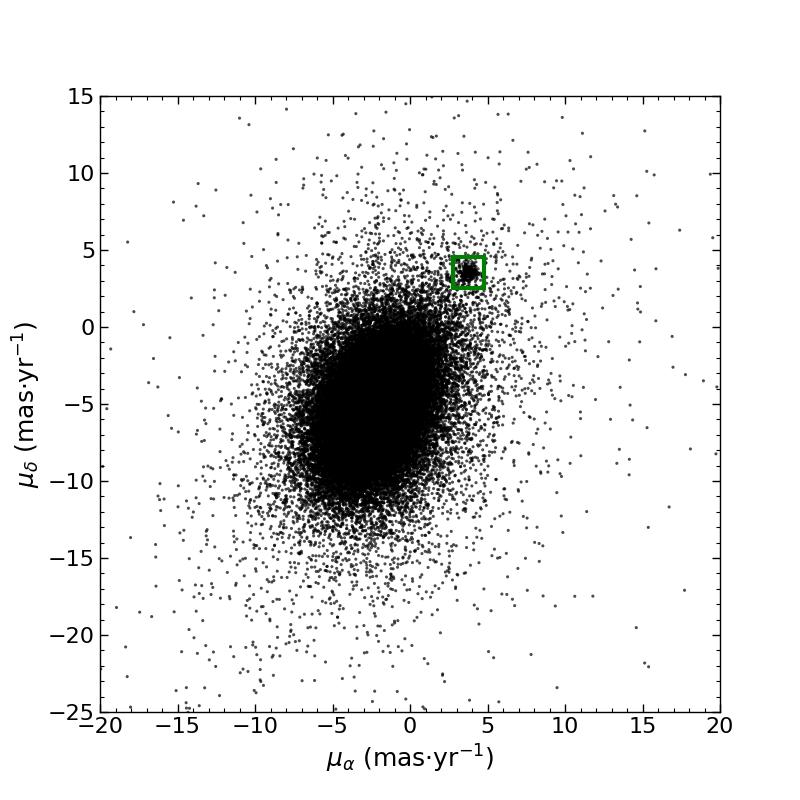}
		\caption{Vector point diagram (VPD) showing the proper motion distribution ($\mu_{\alpha}$ versus $\mu_{\delta}$) for stars within $20^{\prime}$ of the center of NGC~6793. The green square represents the kinematic selection window, defined as the cluster center $\pm 2$~mas~yr$^{-1}$ in both components. This selection criteria isolated 2061 stars, which were used for the subsequent determination of structural parameters.}
		\label{fig-PM}
	\end{figure}

		For the determination of the structural parameters, we followed the methodology of \citet{andrae2023} and \citet{Angelo2025}. The distance of each star from the adopted cluster center (Table 1) has been calculated. Figure~\ref{fig-PM} shows the proper motion distribution ($\mu_{\alpha}$ versus $\mu_{\delta}$) for stars located within $20^{\prime}$ of the cluster center. Stars enclosed within the green square, defined as the cluster center $\pm 2$ mas~yr$^{-1}$ in both proper motion components, were selected to adequately cover the cluster region. This selection resulted in 2061 stars.
	
The structural parameters of the cluster — namely the core radius ($R_c$), tidal radius ($R_t$), central density ($\sigma_0$), and background density ($\sigma_{\mathrm{bg}}$) — were derived by fitting the empirical King (1962) surface density profile to the radial density profile (RDP). The King profile is expressed as:
	
	\begin{equation}
		\begin{split}
			\sigma_{R} = 
			\begin{cases}  
				\sigma_{bg} + k \Bigg[ \frac{1}{\sqrt{1+(R/R_c)^2}} 
				- \frac{1}{\sqrt{1+(R_t/R_c)^2}} \Bigg]^{2}, & R \leq R_t, \\[1ex]
				\sigma_{bg}, & R > R_t,
			\end{cases}
		\end{split}
		\label{eq:king62}
	\end{equation}
	
	where $\sigma(R)$ represents the projected stellar density at radius $R$, and $k$ is given as below.
	
	\begin{equation}
		k = \sigma_o \left[ 1 - \dfrac{1}{\sqrt{1+(R_t/R_c)^2}} \right]^{-2}  
	\end{equation}
	
\begin{figure}
		\centering
		\includegraphics[width=7.5cm]{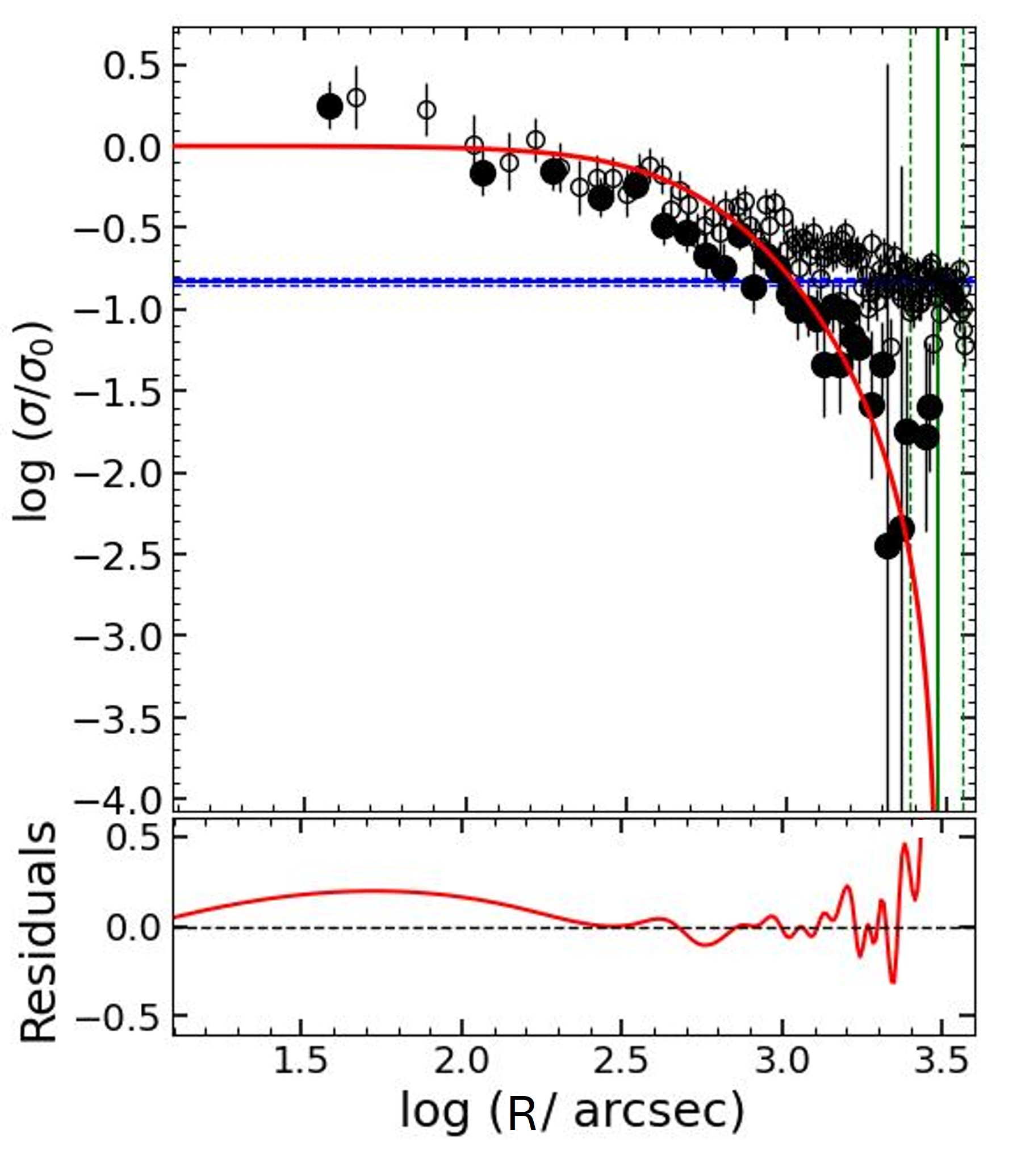}
		\caption{Radial Density Profile (RDP) for NGC 6793. Top panel: the filled and open symbols represent the background-subtracted stellar surface density profiles, normalized and non-normalized, respectively. Error bars are derived from Poisson statistics. The continuous red line represents the best-fit King (1962) profile. The vertical solid and dashed green lines indicate the tidal radius ($R_t$) and its associated uncertainty; and the horizontal solid and dashed blue lines represent the mean background density level and its uncertainty. Bottom panel: the residuals between the observed data and the King model fit. The individual black points in the residual panel represent the $\log(O-C)$ (observed minus calculated) values for each radial bin, while the red line indicates the smoothed residual trend. The dashed black line corresponds to the zero residual level.}
		\label{fig-rdp}
	\end{figure}

	A grid of central coordinates was constructed around the nominal cluster center (Table 1). For each tentative center, the stellar field was divided into concentric annuli of increasing radius. The stellar density in each annulus was computed as $\sigma(R) = N_\ast / A(R)$, where $N_\ast$ is the number of stars within the annulus and $A(R)$ is its area. The background density $\sigma_{\mathrm{bg}}$ and its uncertainty were estimated from the mean stellar density of the outermost annuli, where density fluctuations become approximately constant. The background-subtracted RDP was then fitted using a $\chi^2$ minimization procedure. The adopted center corresponds to the configuration that yields the highest central density and the smallest residuals (Fig.~\ref{fig-rdp}).
	
	The core radius $R_c$ is defined as the radius at which the surface density drops to half of the central value, i.e., $\sigma(R_c) = \sigma_0 / 2$. The tidal radius $R_t$ corresponds to the radius where $\sigma(R_t) = \sigma_{\mathrm{bg}}$. The central density contrast was computed as $\sigma_c = (\sigma_0 / \sigma_{\rm bg}) + 1$ following \citet{Bonatto2005}. Uncertainties in $R_c$ and $R_t$ were estimated from the dispersion of the best-fit parameters, taking into account both the residuals of the King fit and the uncertainties in the radial density bins. For each radial bin, the uncertainty in the stellar density was computed by combining the Poisson error from star counts and the background density uncertainty in quadrature. A bootstrap resampling procedure was additionally applied to account for statistical fluctuations in the stellar density profile.

	The empirical King (1962) model was preferred over dynamical models (e.g., \cite{King1966}; \cite{wilson75}) due to its analytical simplicity and its proven effectiveness in describing open clusters. We obtained the following structural parameters: $\sigma_{0} = 1.07 \pm 0.08$ stars~arcmin$^{-2}$, $\sigma_{\mathrm{bg}} = 0.15 \pm 0.01$ stars~arcmin$^{-2}$, $R_c = 9.53 \pm 2.02$ arcmin, and $R_t = 48.93 \pm 4.40$ arcmin. The goodness-of-fit is characterized by a reduced $\chi^2_\nu = 1.19$, indicating an acceptable fit. To further evaluate the reliability of the King model, we examined the residual distribution (Fig.~\ref{fig-rdp}, bottom panel). The residuals fluctuate around zero within approximately $\pm0.25$ over most radii and reach at most $\sim0.5$ near the outer boundary, confirming a statistically robust representation. Our value of $R_c = 1.62 \pm 0.34$ pc is consistent with \cite{Tarr2022} and H23. Similarly, our tidal radius $R_t = 8.34 \pm 0.75$ pc agrees with \citet{Tarr2022}, \citet{Angelo2025}, and \citet{Tasdemir2025}. For comparison, cluster dimensions reported in degrees by H23 were converted into arcminutes. In contrast, \citet{alm2025} report a significantly larger tidal radius with substantial uncertainty. As emphasized by \citet{Tarr2022}, the determination of tidal radii is intrinsically difficult due to loose cluster structures and field star contamination.

	\renewcommand{\tabcolsep}{1.5mm}
	\renewcommand{\arraystretch}{1.6}
	\begin{table*}[htb]
		\tabularfont
		\caption{Structural parameters of NGC\,6793.}\label{tbl-struct}\vspace*{-2mm}
		\resizebox{1.0\textwidth}{!}{ 
			\begin{tabular}{l c c c  | c c c c c}
				\topline
				
				$\sigma_{0K}$ & 
				$\sigma_{bg}$ & 
				$R_{c}$ &  
				$R_{t}$ & 
				$\sigma_{0K}$ & 
				$\sigma_{bg}$ & 
				$R_{c}$ &  
				$R_{t}$ & 
				Reference \\
				($*\,\prime^{-2}$) & ($*\,\prime^{-2}$) & ($\prime$) & ($\prime$) & ($star\,pc^{-2}$) & ($star\,pc^{-2}$) & (pc) & (pc) & \\
				\midline
				
				1.07$\pm$0.08 &0.15$\pm$0.01 & 9.53$\pm$2.02 & 48.93$\pm$4.40 & 36.82$\pm$0.34 & 5.16$\pm$0.34 & 1.62$\pm$0.34 & 8.34$\pm$0.75 &  This study \\
				\midline
				--&-- & 10.35$\pm$2.11 & 53.28$\pm$2.28 &-- &-- & 1.86$\pm$0.38 & 9.59$\pm$0.41 & 1 \\
				--&-- & 7.52 & 92.17 &-- &-- & 1.28 & 15.67 & 2 \\
				--& --& 3.27$\pm$0.18 & $369.40^{+144.40}_{-169.00}$ & --& --& 0.56$\pm$0.03 & $63.32^{+24.72}_{-28.96}$ & 3 \\
				--& --& 5.29$\pm$0.59 & 38.82$\pm$15.29 &-- &-- & 0.90$\pm$0.10 & 6.60$\pm$2.60 & 4 \\
				89.63$\pm$5.88& 62.86$\pm$0.75& 1.59$\pm$0.13 & 40.90$\pm$2.20 &  2972.03$\pm$194.97 &2084.36$\pm$24.87 & 0.28$\pm$0.02  & 7.11$\pm$0.38 & 5 \\			
				\hline
				
			\end{tabular}
		}
		
		\tablenotes{($*\,\prime^{-2}$) and ($*\,pc^{-2}$) in Cols.~1--2 and 5--6 }represent stars~arcmin$^{-2}$ and stars~pc$^{-2}$, respectively. For the conversion of arcmin to pc, we adopt the Gaia distance (Row 1 of Table~6). The first row lists our results; the subsequent rows are literature data. References: 1: \cite{Tarr2022}, 2: \cite{hunt2023}, 3: \cite{alm2025}, 4: \cite{Angelo2025}, 5: \cite{Tasdemir2025}. All references provide only $R_{c}$ and $R_{t}$ values, except for this study and \cite{Tasdemir2025}. The goodness-of-fit of the RDP model for this study (first row) is characterized by a reduced $\chi^{2}_{\nu}=1.19$.
		
	\end{table*}
	
	To assess the robustness of the derived parameters, the King fitting procedure was repeated using stellar samples extracted within radii of $30^{\prime}, 45^{\prime}, 75^{\prime}, 90^{\prime},$ and $120^{\prime}$ (Table~\ref{tab:struct_sensitivity}). The background density $\sigma_{\mathrm{bg}}$ and central density $\sigma_0$ remain remarkably stable across the full range of extraction radii, with values ranging between $0.14-0.19$ and $1.01-1.28$ stars~arcmin$^{-2}$, respectively. While $R_c$ varies between $7^{\prime}.07$ and $13^{\prime}.75$ and $R_t$ ranges from $44^{\prime}.98$ to $53^{\prime}.32$, these fluctuations are minimal and show no systematic trends. This confirms that the adopted extraction radius of $60^{\prime}$ provides reliable structural parameters.
	
	\renewcommand{\tabcolsep}{1mm}
	\renewcommand{\arraystretch}{1.2}
	\begin{table}[htb]
		\tabularfont
		\caption{Sensitivity of structural parameters to the spatial variations in the background selections. $R$ in the first column represents the radius chosen for data. All parameters are given in arcmin.}
		\resizebox{\columnwidth}{!}{%
				\begin{tabular}{c c c c c}
					\hline
					$R$ & $\sigma_{0K}$ & $\sigma_{bg}$ & $R_{c}$ & $R_{t}$\\
					\hline		
					
					30 & 1.13$\pm$0.17 & 0.19 $\pm$0.02 & 9.23$\pm$3.29 & 47.10$\pm$11.16 \\		
					45 & 1.28$\pm$0.07 & 0.18 $\pm$0.01 & 7.07$\pm$1.42 & 44.98$\pm$6.40 \\
					60 & 1.07$\pm$0.08 &  0.15 $\pm$0.01 & 9.53$\pm$2.02 & 48.93$\pm$4.40 \\
					75 & 1.22$\pm$0.09 &  0.16 $\pm$0.01 & 9.97$\pm$4.11 & 51.89$\pm$5.83 \\
					90 & 1.04$\pm$0.02 &  0.16 $\pm$0.01 & 11.86$\pm$3.21 & 48.57$\pm$5.03 \\
					120 & 1.01$\pm$0.07 &  0.14$\pm$0.01 & 13.75$\pm$2.84 & 53.32$\pm$17.93 \\
					\hline
					
		\end{tabular}}
		\label{tab:struct_sensitivity}
	\end{table}

	\subsection{Membership Selection}
	
	For membership selection, we utilized the KMeans method implemented in pyUPMASK \citep{per21}, a Python-based evolution of the UPMASK (Unsupervised Photometric Membership Assignment in Stellar Clusters) framework \citep{kron2014}. Our sample (682 stars) was processed using five primary parameters-position $(\alpha, \delta)$, proper motions $(\mu_{\alpha} , \mu_{\delta})$, and parallax $(\varpi)$ to assign a membership probability to each source. We configured the clustering engine with a target of 25 stars per cluster and a maximum limit of 20 clusters, resulting in a partitioning of the data into approximately 20 groups per iteration. To ensure the robustness of the membership assignments, we executed 30 outer loop runs without a fixed random seed to allow for stochastic exploration of the parameter space. During each iteration, the pipeline performed outlier removal and used Principal Component Analysis (PCA) for dimensionality reduction to better isolate the cluster signal from background noise. The final membership probabilities were then refined using a Gaussian-Uniform Mixture Model (GUMM) and Kernel Density Estimation (KDE) to distinguish true members from field stars, with the entire process optimized via NumPy multithreading for efficient computation.

\begin{figure}[!t]
		\centering
		\includegraphics[width=0.35\textwidth]{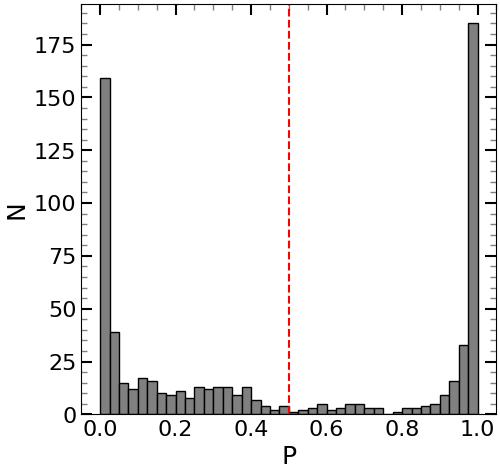}
		\caption{Histogram of membership probabilities for 682 stars computed using \texttt{pyUPMASK}. The red dashed line indicates the adopted membership threshold of $P = 0.5$, used to select high-confidence cluster members.}
		\label{fig-probs}
	\end{figure}

	As a result, we determined the membership probabilities for stars within the cluster region ($R_{t} < 48'.93$). Fig~\ref{fig-probs} presents the histogram of these probabilities. In the literature, the most likely members are frequently defined as those with probabilities $P > 0.5$ \citep[e.g.,][]{cantat2018, carr2019, dia21, hunt2023, alm2025}. To assess the impact of this threshold on our results, we tested several cutoff values: $P = 0.5, 0.6, 0.7, 0.8,$ and $0.9$. Table~\ref{tab:membership_sensitivity} summarizes the median values of $\varpi$, $\mu_{\alpha}^*$, and $\mu_{\delta}$, along with the number of identified members (N) for each threshold. We found that the median astrometric parameters vary only minimally across the different cutoffs, with ranges of 0.003~mas for $\varpi$, 0.011~mas~yr$^{-1}$ for $\mu_{\alpha}$, and 0.009~mas~yr$^{-1}$ for $\mu_{\delta}$. While N decreases from 291 to 243 stars as the probability cutoff increases to 0.9, the stability of the astrometric medians suggests that a more inclusive threshold does not introduce significant field contamination. Although the probability distribution rises markedly around $P = 0.9$ (Fig~\ref{fig-probs}), we adopted $P > 0.5$ as our final cutoff to ensure a complete sample of 291 member stars. Stars with probabilities below this threshold were excluded to minimize field contamination while maintaining a reliable and representative membership list.
	
	\renewcommand{\tabcolsep}{1mm}
	\renewcommand{\arraystretch}{1.2}
	\begin{table}[htb]
		\tabularfont
		\caption{Sensitivity of the membership selection to probability (P) thresholds for 682 stars. Median astrometric parameters and number of members (N) are listed for each threshold.}
		\resizebox{\columnwidth}{!}{%
			\begin{tabular}{c c c c c}
				\hline
				P-threshold & N & $\varpi$ (mas) & $\mu_{\alpha}$ (mas yr$^{-1}$) & $\mu_{\delta}$ (mas yr$^{-1}$) \\
				\hline
				0.5 & 291 & 1.705 $\pm$0.057 & 3.786$\pm$0.051 & 3.538$\pm$0.058 \\
				0.6 & 280 &  1.704 $\pm$0.054 & 3.786$\pm$0.048 & 3.543$\pm$0.054 \\
				0.7 & 265 & 1.705$\pm$0.049 & 3.794$\pm$0.045 & 3.546$\pm$0.050 \\
				0.8 & 258 & 1.707$\pm$0.048 & 3.797$\pm$0.043 & 3.547$\pm$0.049 \\
				0.9 & 243 & 1.705$\pm$0.046 & 3.797$\pm$0.041 & 3.547$\pm$0.044 \\
				\hline
		\end{tabular}}
		\label{tab:membership_sensitivity}
	\end{table}

\begin{figure*}[!t]
		\centering
		\includegraphics[width=1.0\textwidth]{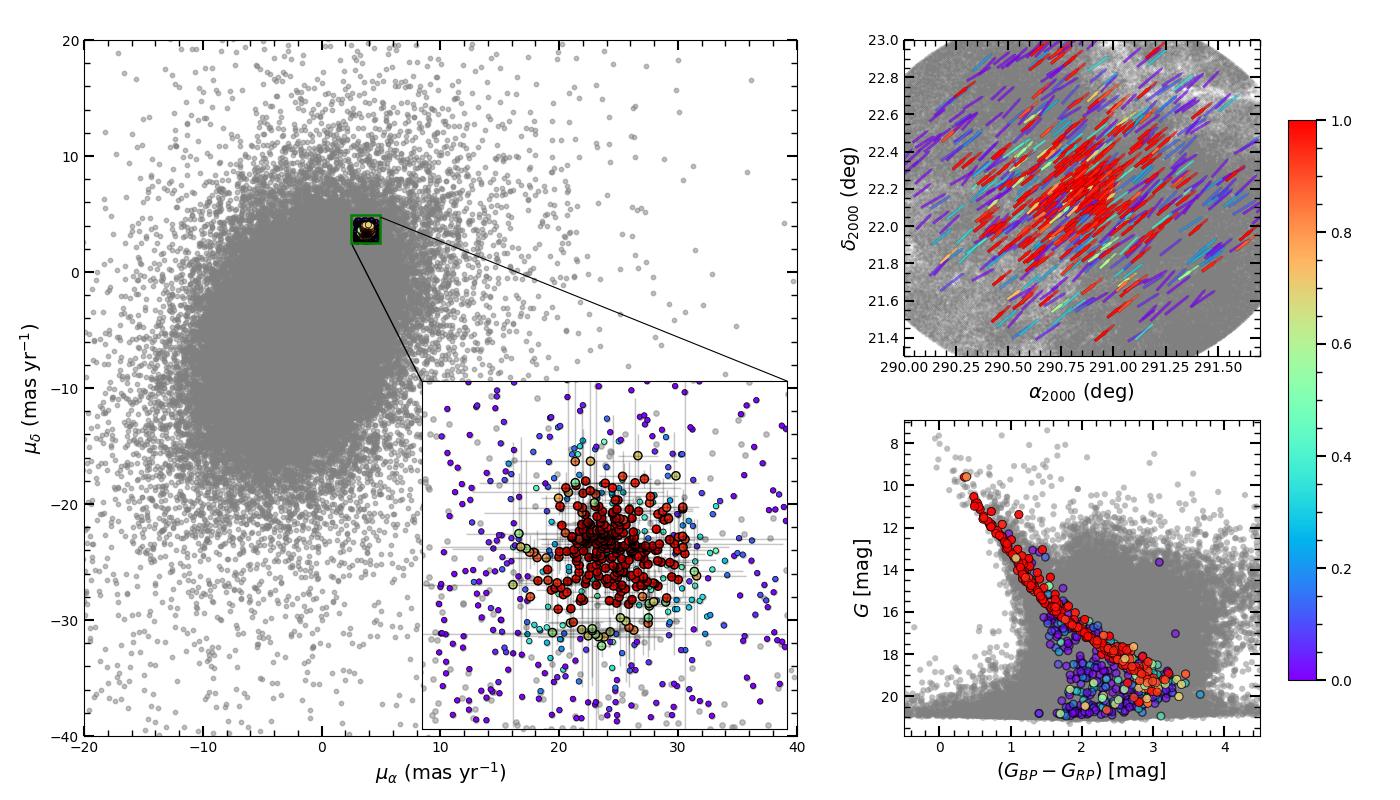}
		\caption{Vector-point diagram (VPD; left panel), proper-motion velocity vectors (top-right panel), and 
			$G$ vs.\ $(G_{BP}-G_{RP})$ CMD (bottom-right panel) of NGC~6793. Candidate members (682 stars) are 
			colour-coded according to their membership probability (colour bar on the far right), while field stars 
			are shown in gray (633\,937 stars). The VPD inset shows uncertainties in $\mu_{\alpha}$ and $\mu_{\delta}$ as horizontal and vertical black bars, respectively, while in the CMD, uncertainties in $G$ and $G_{BP}-G_{RP}$ are indicated by vertical and horizontal black error bars. The 291 high-probability members ($P > 0.5$) are 
			highlighted with larger symbols in the proper-motion velocity-vector diagram for improved visibility. Proper motions are plotted in the Gaia reference frame using $(\mu_{\alpha},\,\mu_{\delta})$, with the correct axis orientation and vector scaling.}	
		\label{fig-vpd-cmd}
	\end{figure*}

	Median proper motion components of 291 members of NGC~6793 are determined as $\mu_{\alpha} = 3.786 \pm 0.051$~mas~yr$^{-1}$ and $\mu_{\delta} = 3.538 \pm 0.058$~mas~yr$^{-1}$, with respective ranges of (3.225, 4.281)~mas~yr$^{-1}$ and (2.978, 4.071) mas yr$^{-1}$. The median parallax has been measured as $\varpi = 1.705 \pm 0.057$~mas, with a range of (1.314, 1.992)~mas. Vector-point diagram (VPD), proper-motion velocity vectors, and the $G_{BP}-G_{RP}$ CMD of NGC 6793 are presented in Fig.~\ref{fig-vpd-cmd}. Candidate members (682 stars) are colour-coded according to their membership probability, while field stars are shown as gray points. In the VPD (Figure~\ref{fig-vpd-cmd}, left panel), the 291 high-probability members ($P > 0.5$) form a compact concentration rather than a sparse distribution; these same stars trace a clear main sequence in the CMD (Figure~\ref{fig-vpd-cmd}, bottom-right panel). Uncertainties in $\mu_{\alpha}$ and $\mu_{\delta}$ are indicated by gray horizontal and vertical bars in the VPD inset. Its median Gaia astrometric values are listed in Table~\ref{tbl-propermotion}, alongside the literature estimates. While a simple inverse-parallax relation yields a distance of $587 \pm 19$~pc, we also applied the Bayesian estimation method of \citet{Bailer2018}\footnote{\url{https://bailer-jones.www3.mpia.de/gedr3_distances.html}} to account for parallax uncertainties and the non-linearity of the distance–parallax relation. This yields a consistent distance of $586 \pm 10$~pc. Within the uncertainties, our median values for these 291 members are compatible with those reported by \citet{cantat2020}, \citet{dia21}, and H23. The estimated median equatorial coordinates are $(\alpha, \delta) = (290^{\circ}.807, +22^{\circ}.152)$, corresponding to Galactic coordinates $(l, b) = (56^{\circ}.172, 3^{\circ}.329)$, both of which are in good concordance with the values in Table~\ref{tab:coords}.
	
     The SPM survey of NGC~6793 includes a sample of 1283 stars with CCD $UBV(RI)_{KC}$ photometric data. To integrate this with the Gaia DR3 catalog, we cross-matched the equatorial coordinates $(\alpha, \delta)$ of the 291 high-probability members ($R_{t} < 48'.93$) with the SPM dataset using a 1~arcsec matching radius. To ensure high-quality photometric fitting, we applied a quality cut requiring the uncertainty in the $(U-B)$ color index to be less than 0.1~mag. This procedure yielded a final sample of 34 probable members with reliable CCD $UBV(RI)_{KC}$ photometry, which is utilized in the subsequent analyses.

	\renewcommand{\tabcolsep}{1mm}
	\renewcommand{\arraystretch}{1.2}
	\begin{table*}[htb]
		\tabularfont
		\centering
		\caption{The median proper motion components and trigonometric parallax/distance of the 291 likely members of NGC~6793 derived in this study (top row). Literature values are given in the following rows.}
		\label{tbl-propermotion}
		\begin{tabular}{lcccccc}
			\topline
			$\mu_{\alpha}$ & $\mu_{\delta}$ & $\varpi$ & $d$ & N & Reference\\
			(mas/yr) & (mas/yr) & (mas) & (pc) &&  \\ 
			\midline
			
			3.786$\pm$0.051 & 3.538$\pm$0.058 & 1.705$\pm$0.057 & 586$\pm$10 & 291 &This study \\
			\midline
			3.800$\pm$0.188 & 3.536$\pm$0.201 & 1.671$\pm$0.058 & 586 & 262 & 1 \\
			3.795$\pm$0.186 & 3.544$\pm$0.177 & 1.669$\pm$0.055 & 628 & 185 & 2\\
			3.778$\pm$0.194 & 3.569$\pm$0.237 & 1.667$\pm$0.055 & 589$\pm$7 & 187 & 3\\
			\hline
		\end{tabular}
		\tablenotes{1: \cite{hunt2023}, 2: \cite{cantat2020}, 3: \cite{dia21}
		}
	\end{table*}

	\section{Astrophysical Parameters of NGC 6793}\label{sec-astro}
	
	\subsection{Reddening}\label{sec:redd}
\begin{figure}[!t]
		\centering
		\includegraphics[width=0.7\linewidth]{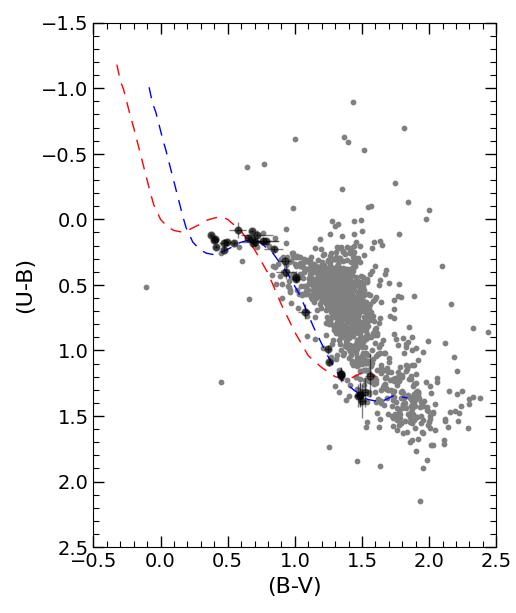}
		\caption{The CC diagram for 34 members (filled black dots). Dashed red and blue lines show the intrinsic and reddened S13 dwarf sequences. Gray points denote the cluster stars (N$=$1283) with CCD~$UBVRI$.}
		\label{fig-4-CCVUB}
	\end{figure}

	The $(U-B)$ vs. $(B-V)$ colour-colour diagram (CC) for the 34 eligible members of NGC~6793 with available CCD $UBV(RI)_{KC}$ photometry (filled circles) is presented in Figure~\ref{fig-4-CCVUB}. The interstellar reddening $E(B-V)$ has been obtained by shifting the intrinsic-color sequence of \citet{sung13} until a best fit to the member stars has been achieved. This shift accounted for a $B-V$ displacement of $E(B-V)$ and a $U-B$ displacement following the relation $E(U-B) = 0.72E(B-V) + 0.025E(B-V)^2$. We obtained a mean reddening of $E(B-V) = 0.24 \pm 0.05$~mag, which corresponds to $E(U-B) = 0.17 \pm 0.03$, $E(V-I) = 0.30 \pm 0.05$, $E(R-I) = 0.17 \pm 0.03$, and $E(G_{BP}-G_{RP}) = 0.31 \pm 0.05$. The cluster members exhibit a tight distribution around the reddened colour sequence, confirming the reliability of the fit.

	We quantitatively assessed how small-number statistics, the colour–colour distribution, photometric errors, and unresolved binaries affect our reddening estimate. Using Monte Carlo simulations based on the 34 members with CCD $UBV(RI)_{KC}$ photometry, we perturbed each star’s $(B-V)$ values within their measured uncertainties and repeated the reddening determination over 5000 realizations. The simulation yielded a mean distribution of $E(B-V) = 0.280$~mag, identifying a systematic bias of $+0.034$~mag. By correcting for this bias and propagating the random uncertainty ($\sigma_{ran} = \pm0.004$~mag), we derived a bias-corrected simulation result of $E(B-V) = 0.246 \pm 0.034$~mag. This result is in near-perfect agreement with our initial CC-diagram estimate of $0.24$~mag.
	
	The small scatter confirms that the measured photometric uncertainties (ranging from $0.007$–$0.060$ in $B-V$ and $0.006$–$0.083$ in $U-B$) contribute primarily to random dispersion in the CC diagram rather than significant systematic offsets. Furthermore, while unresolved binaries can introduce a small positive bias in $E(B-V)$, typically $\lesssim 0.01$–$0.03$~mag \citep{Fernie1961}, this effect primarily impacts stars earlier than B9 ($T_{eff} \gtrsim 10,500$~K). Effective temperatures ($T_{eff}$) available for our members from Gaia DR3 (\texttt{GSP-Spec}, \texttt{ESP-HS}) and SPM observations confirm that the members of NGC~6793 are cooler than B9 stars, with $T_{eff}$ values peaking below $9800$~K. Consequently, the impact of unresolved binaries on the derived reddening is considered negligible. By following these checks, we adopt $E(B-V) = 0.24 \pm 0.05$~mag for NGC~6793, which is consistent with both our simulation result and the literature estimates (0.20–0.40~mag; see Table~\ref{tab:litcompbig}).

	\subsection{Metal abundance}\label{sec:met}
	
	No spectroscopic metallicity data for NGC~6793 are found in the literature or within the LAMOST, GALAH, or APOGEE surveys for the 291 identified members. Furthermore, none of these stars have metallicity estimates in the Gaia GSP-Spec module. Consequently, we applied the photometric metallicity calibration of \citet{karatas2006}, $[M/H] = +0.09 - 3.01\delta_{0.6} - 16.58\delta_{0.6}^2$.  Here $\delta_{0.6}$ means the normalized ultraviolet excess, and is obtained from the the selected F type members on the main sequence in the CC diagram (Fig.~7). See \citet{karatas2006}'work for the details. This yielded a mean metallicity of $[Fe/H] = +0.20$. We assume $[Fe/H]=[M/H]$. Adopting the solar metallicity $Z_\odot = 0.0152$ \citep{Caffau2011} and the standard relation $[Fe/H] = \log_{10}(Z/Z_\odot)$, this corresponds to $Z = 0.024$. This value is in good agreement with previous photometric studies, such as $Z = 0.026$ \citep{dia21}, $Z = 0.020$ \citep{alm2023}, and $Z = 0.025$ \citep[A25;][]{cantat2020}.
	
	The derived cluster metallicity ($[Fe/H] \approx +0.20$) is relatively high for a cluster at a peri-galacticon of 7.37~kpc, where values of $[Fe/H] \approx +0.1$ to $+0.2$~dex are typically expected. However, similar elevated metallicities are observed in other open clusters at comparable Galactic distances, such as Praesepe ($[Fe/H] = +0.27 \pm 0.10$, $R_{GC} \approx 7.6$~kpc; \citet{pace2008}) and NGC~6231 ($[Fe/H] = +0.26 \pm 0.10$, $R_{GC} \approx 6.7$~kpc) \citep{pau2010}. These examples suggest that such metallicities at these Galactic radii are not unprecedented and may result from localized enrichment or radial migration. In the absence of high-quality spectroscopy, our calibrated estimate remains a reliable and consistent constraint for NGC~6793, pending future spectroscopic confirmation.

	\subsection{Age and Distance}\label{sec:bayes}

	NGC 6793 possesses only a sparse population of evolved stars near the Main Sequence Turn-Off (MSTO) (Fig.~10), and the low-mass pre–main-sequence "turn-on" is incomplete in Gaia photometry. Furthermore, gyrochronology, which derives ages from stellar rotation, cannot be applied here as the required rotation periods for cluster members are currently unavailable in the literature. Consequently, to overcome the lack of a well-defined turn-off or rotation-based age, we performed a Bayesian grid-based isochrone fitting following the formalism of \citet{nay06}.This approach utilizes the entire main sequence (291 Gaia-selected members) to provide a statistically robust age and distance. The explored parameter space covered $\log(t/\text{yr}) = 7.0\text{--}10.0$ and $\mu = 7.0\text{--}10.0$ mag, both in steps of 0.02. The posterior probability distribution, constructed from a $\chi^2$-based likelihood comparison between observed and model CMDs, was marginalized over each parameter to derive the 68\% credible intervals. We employed the  isochrones \citep{bre12} (PARSEC isochrones) with fixed metallicity ($Z = 0.024$) and reddening ($E(B-V) = 0.24$).
	
\begin{figure*}[!t]
		\centering
		\includegraphics[width=12cm]{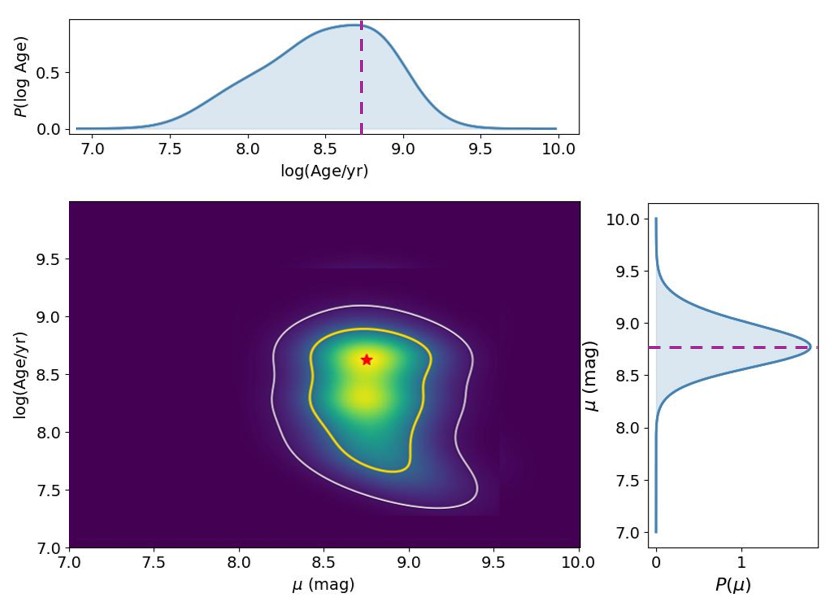}
		\caption{
			Joint and marginalized posterior probability distributions for the cluster age and distance modulus $\mu$ from Bayesian isochrone fitting, assuming $Z=0.024$ and $E(B-V)=0.24$. 
			The central panel shows the joint posterior $P(\log(Age/yr),\mu)$, with the red star marking the MAP solution ($\log(Age/yr)_{MAP}=8.74$, $\mu_{MAP}=8.77$~mag). 
			Gold and white contours enclose the 68\% and 95\% credible regions, respectively. The colour scale corresponds to posterior probability density, with yellow representing the highest and blue the lowest values. Top and right panels display the marginalized posteriors for $\log(Age/yr)$ and $\mu$, with dashed red lines indicating the MAP values.
		}
		\label{fig:beyes_age}
	\end{figure*}

	To ensure the robustness of the solution, we performed the fitting using the full membership catalog and then repeated the analysis restricted to stars below the nominal turn-off region; the resulting parameters remained consistent within $1\sigma$, confirming that our age is well-constrained by the MS slope despite the scarcity of evolved stars.The maximum–a–posteriori (MAP) solution yields $\log(t/\text{yr}) = 8.74 \pm 0.08$ ($t = 550 \pm 100$ Myr) and $\mu = 8.77 \pm 0.22$ mag ($d = 575 \pm 58$ pc), consistent within errors with the distances given in Tables \ref{tbl-propermotion} and \ref{tab:litcompbig}. This result, represented by the red star in the joint posterior panel of Fig.~\ref{fig:beyes_age}, lies near the peak of the probability distribution. A mild secondary maximum in the joint posterior reflects the usual age–distance degeneracy, though the primary peak remains the most probable. The relatively broad age uncertainty primarily arises from the small number of evolved stars and the limited evolutionary leverage along the main sequence \citep[e.g.,][]{Bon2007, bossini2019}.

	\subsection{Simultaneous Multi-CMD Isochrone Fitting and Parameter Optimization}\label{sec:simul}
	
\begin{figure*}[!t]
		\centering
		\includegraphics[width=14cm]{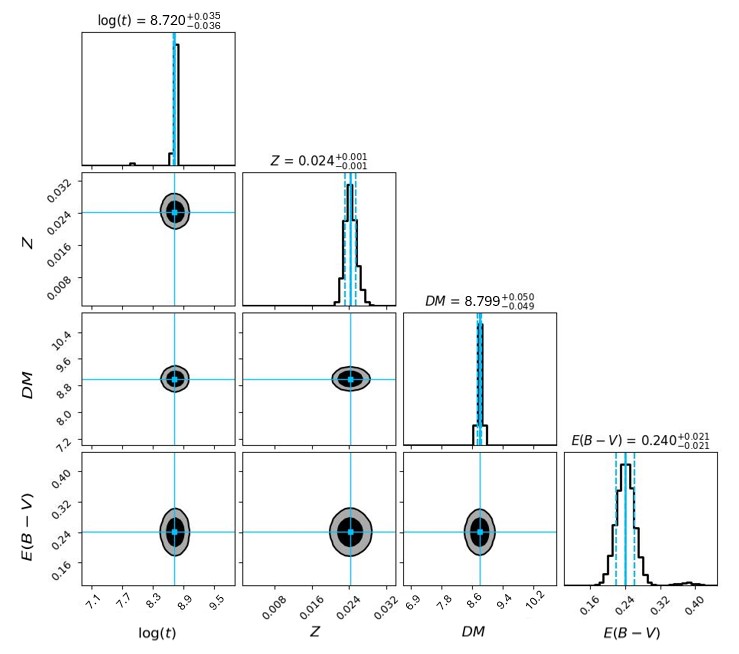}
		\caption{Corner plot illustrating the marginalized posterior probability distributions for the cluster parameters, derived from a Bayesian grid-based isochrone fitting. The diagonal panels show the 1D histograms for age ($\log(t)$), metallicity ($Z$), distance modulus ($DM$), and reddening ($E(B-V)$), where the solid cyan lines indicate the median values and dashed lines represent the $1\sigma$ (16th and 84th) percentiles. The off-diagonal plots display 2D contour maps representing the $0.68$ and $0.95$ confidence levels. The samples were generated by weighting the grid search $\chi^2$ results with a likelihood function $P \propto \exp(-\chi^2/2)$ and applying a Gaussian jitter to account for the discrete grid steps.
		}
		\label{fig:corner}
	\end{figure*}

	A simultaneous $\chi^{2}$ optimization has been performed to determine the best-fit astrophysical parameters of the cluster for 34 members with CCD $UBV(RI)_{KC}$ photometry. This was achieved using the simul\_fit code\footnote{\url{https://github.com/speletry/simul\_fit.py}}, a Python tool designed for the simultaneous multi-filter isochrone fitting of stellar clusters. The fit was carried out jointly across the CMDs $V,(B-V)$, $V,(U-B)$, $V,(V-I)$, and $V,(R-I)$, where each star’s observed magnitude and colour, along with their photometric uncertainties, contribute to the global $\chi^{2}$ statistic. The parameters $Z$, $E(B-V)$, $\mu$, and $\log(Age)$ are optimized simultaneously. The resulting corner plot (Fig. \ref{fig:corner}) shows a best-fit solution of $Z=0.024\pm0.001$, $\log(Age/\text{yr})=8.72\pm0.04$, $\mu=8.80\pm0.05$, and $E(B-V)=0.24\pm0.02$, with a global reduced $\chi^{2}=0.94$. This indicates excellent agreement between the observed and synthetic CMDs. The simultaneous multi-CMD fitting procedure also accounts for the interdependence of the parameters, revealing modest correlations between $E(B-V)$ and $\mu$, and between $Z$ and $\log(Age)$, as expected for partially degenerate parameters. The full covariance matrix of the fitted parameters is:
	
	\[
	\begin{pmatrix}
		0.014979 & -0.000033 & -0.000057 & -0.002554 \\
		-0.000033 &  0.000002 &  0.000007 &  0.000004 \\
		-0.000057 &  0.000007 &  0.002736 & -0.000036 \\
		-0.002554 &  0.000004 & -0.000036 &  0.000848
	\end{pmatrix}
	\]
	
\begin{figure*}[!t]
		\centering
		\includegraphics[width=0.36\textwidth, height=6cm]{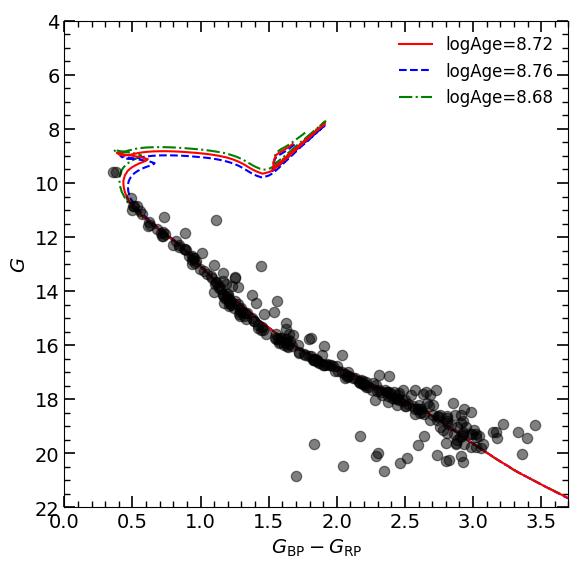}\\
		\includegraphics[width=0.36\textwidth, height=6cm]{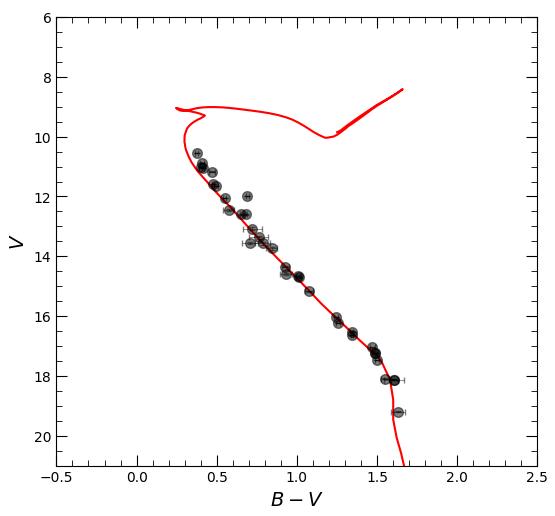}\hspace{1mm}
		\includegraphics[width=0.36\textwidth, height=6cm]{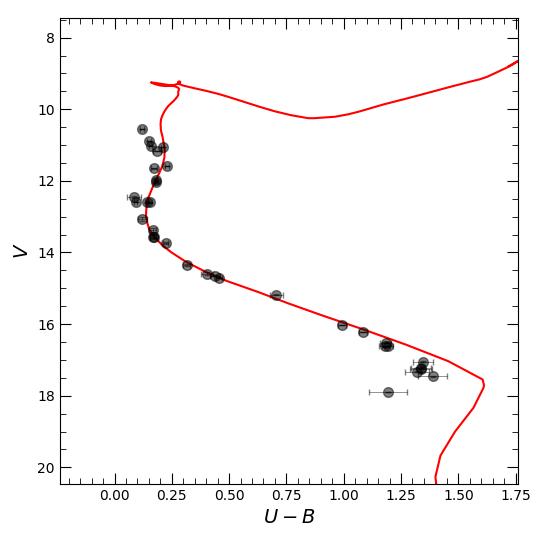}\hspace{1mm}
		\includegraphics[width=0.36\textwidth, height=6cm]{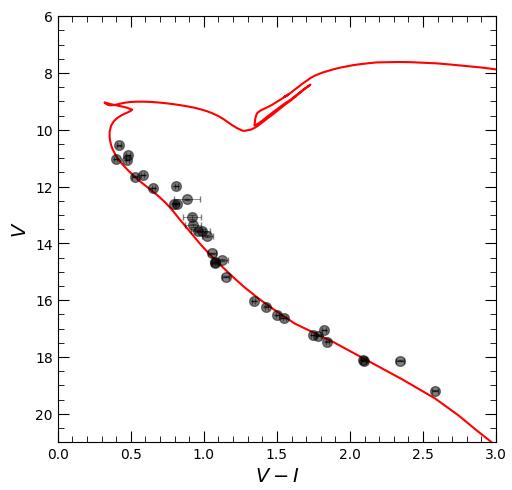}\hspace{1mm}
		\includegraphics[width=0.36\textwidth, height=6cm]{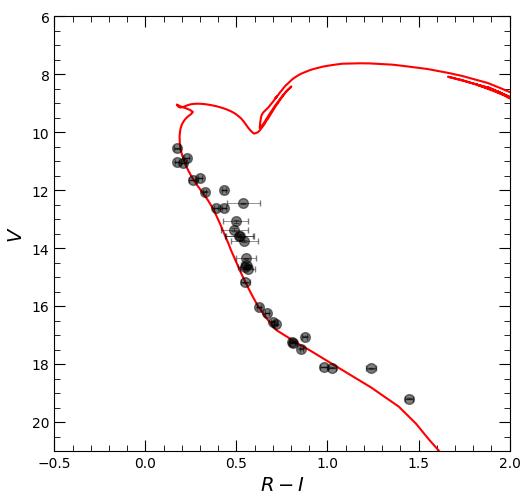}\hspace{1mm}
		
		\caption{
			The CMDs  of NGC\,6793: $G$ vs.\ $(G_{BP} - G_{RP})$ (291 members), 
			$V$ vs.\ $(B - V)$ (34 members), $V$ vs.\ $(U - B)$ (34 members), 
			$V$ vs.\ $(V - I)$ (33 members), and $V$ vs.\ $(R - I)$ (33 members).
			Cluster members are displayed as black open circles together with their uncertainties. 
			The solid red lines represent  the best-fitting PARSEC isochrones in five CMDs.
			The two isochrones of $\log(Age/yr) = 8.68$ and $8.76$ (blue/green dashed lines) in the Gaia CMD have been shown to provide a means for appreciating the uncertainties.}
		\label{fig:cmdubvri}
	\end{figure*}

	The best-fit isochrones are compared to the CMDs in Fig. \ref{fig:cmdubvri}. For the Gaia $G, (G_{BP}-G_{RP})$ CMD, age uncertainties of $\pm0.04$ dex are indicated by dashed lines. The formal $1\sigma$ uncertainties and parameter covariances were derived from the $\chi^{2}$ grid using likelihood weights of the form $P \propto \exp(-\Delta\chi^{2}/2)$. This provides a statistically consistent estimate of the parameter correlations and is equivalent to a Bayesian posterior-weighted covariance analysis. No multimodal behavior was detected in the posterior surface.

	The simultaneously fitted parameters are in excellent agreement with the values independently derived in Sections~\ref{sec:redd}, \ref{sec:met}, and \ref{sec:bayes} ($E(B-V)=0.24\pm0.05$, $Z=0.024$, $\log(Age/\text{yr})=8.74\pm0.08$, $\mu=8.77\pm0.22$). As illustrated in Fig.~\ref{fig:cmdubvri}, these parameters provide a superior representation of the cluster sequences across all colour-magnitude planes. Consequently, we adopt these results as the final cluster parameters, as they demonstrate the robustness and internal consistency of the multi-filter fitting procedure. The fundamental astrophysical parameters of NGC~6793 are summarized in Table~\ref{tbl-astro-lit}, along with a comparison to previous literature values.

	\begin{table*}[htb]
		\tabularfont
		\centering
		\caption{Literature comparison for NGC~6793.}
		\label{tab:litcompbig}
		\small
		\label{tbl-astro-lit}
			\begin{tabular}{lccccccll}
				\topline
				$E(B-V)$ & $\mu$ & $d$~(pc) & $Z$ & $[Fe/H]$ & Age~(Myr) & N & Photometry & Reference \\
				\midline
				0.24$\pm$0.02 & 8.80$\pm$0.05 & 575$\pm$58 & 0.024$\pm$0.001 & +0.20$\pm$0.02 & 525$\pm$51 & 34 &  CCD UBVRI & This paper \\
				0.31$\pm$0.08 & 8.70$\pm$0.20 & -- & 0.025$\pm$0.001 & +0.22$\pm$0.017 & 447$\pm$206 & -- & Gaia DR3 & 1 \\
				0.33$\pm$0.02 & -- & 589$\pm$7 & -- & -- & 458$\pm$73 & -- & Gaia DR3 & 2 \\		
				-- & -- & -- & -- & -- & 309 & 242 & Gaia DR3 & 3 \\	
				0.26$\pm$0.03 & 9.51$\pm$0.07 & 585$\pm$19 & 0.019$\pm$0.002 &  +0.10$\pm$0.05 & 650$\pm$50 &  & Gaia DR3 & 4 \\
				0.33 & -- & 586 & -- & -- & 136 & 439 & Gaia DR3 & 5 \\	
				0.24 & -- & 628 & -- & -- & 309 & 205 & Gaia DR3 & 6 \\	
				-- & -- & -- & -- & -- & 501 & 78 & Gaia DR3 & 7 \\			
				-- & -- & -- & -- & -- & 495 & -- & Gaia DR3 & 8 \\			
				-- & -- & 618 & -- & -- & 309 & 242 & Gaia EDR3 & 9 \\					
				0.45 & -- & -- & -- & -- & 71 & 190 & Gaia EDR3 & 10 \\																
				-- & 8.99 & 628 & -- & -- & 309 & 179 & Gaia EDR3 & 11 \\																
				0.27 & 8.65 & -- & 0.019  & 0.10 & 603 &  & Gaia DR2 & 12 \\																
				0.33$\pm$0.02 & -- & 589$\pm$7 & 0.026$\pm$0.005 & 0.23$\pm$0.08 & 458$\pm$73 & 187 & Gaia DR2 & 13 \\																		
				-- & -- & 589 & -- & -- & -- & 190 & Gaia DR2 & 14 \\										
				0.22$\pm$0.23 & -- & -- & 0.015 & 0.00 & 573 & -- & Gaia DR2 & 15 \\																									
				0.31 & -- & 724 & -- & -- & 495 & -- & 2MASS & 16 \\											
				\hline
			\end{tabular}
			
			\tablenotes{References: 1: \cite{Angelo2025}, 2: \cite{alm2025}, 3: \cite{Xu2025}, 4: \cite{Tasdemir2025}, 5: \cite{Hunt2024}, 6: \cite{Donada2023}, 7: \cite{Liu2023}, 8: \cite{Just2023}, 9: \cite{Tarr2022}, 10: \cite{He2022}, 11: \cite{Poggio2021}, 12: \cite{jad21}, 13: \cite{dia21}, 14: \cite{cantat2020}, 15: \cite{bossini2019}, 16: \cite{Kharchenko2016}.}
	\end{table*}
	
	
	\section{Kinematics and Orbital Parameters of NGC 6793}\label{sec-kinematics}
	
\begin{figure}[!t]
		\centering
		\includegraphics[width=0.42\textwidth]{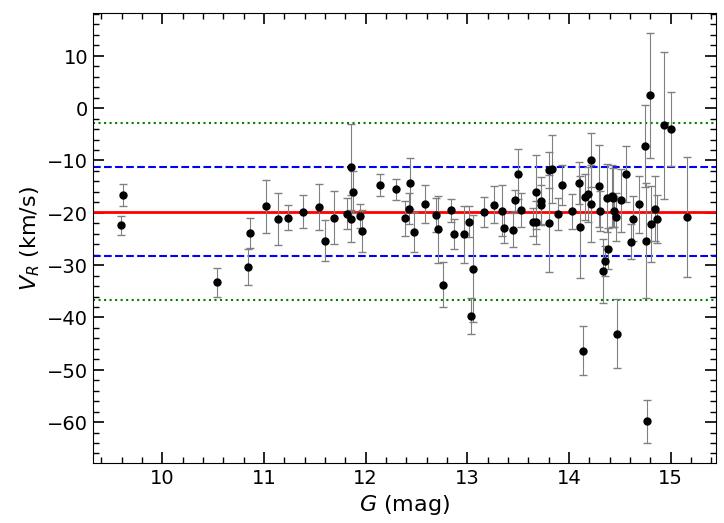}\\
		\caption{
			Radial velocity versus Gaia $G$-band magnitude for the 86 members of NGC~6793.
			The solid red line denotes  the median radial velocity ($-19.78\pm2.47$~km\,s$^{-1}$). 
			The blue dashed and the green dotted  lines indicate the $\pm1\sigma$ and $\pm2\sigma$ limits.}
		
		\label{fig:rv}
	\end{figure}

	Following the formalisms of \citet{joh87}, the heliocentric velocity components ($U, V, W$) of NGC 6793 has been computed in a right-handed coordinate system using its radial velocity ($V_{R}$), proper motion components, and the photometric distance (Section~\ref{sec:bayes}). From the relationship between $V_{R}$ and $G$-magnitude for the cluster members (Fig.~\ref{fig:rv}), the median radial velocity has been determined to be $-19.78 \pm 2.47$~km~s$^{-1}$, based on 86 members within the $\pm 2\sigma$ range. Within the uncertainties this value is consistent with the result reported by \citet{Soubiran2019} ($-19.56 \pm 1.40$~km~s$^{-1}$), \citet{Tasdemir2025} ($-20.02 \pm 0.76$~km\,s$^{-1}$), H23 ($-20.02 \pm 7.22$~km~s$^{-1}$) ,and \citet{dia21} ($-25.68 \pm 4.71$~km~s$^{-1}$).
	
	The heliocentric velocities are corrected to the Local Standard of Rest (LSR) by adopting the solar peculiar motion $(U_{\odot}, V_{\odot}, W_{\odot}) = (11.10, 12.24, 7.25)$~km,s$^{-1}$ \citep{sch10}. Additionally, galactic differential rotation corrections \citep{mih1981} of $\Delta U = 13.05$~km,s$^{-1}$ and $\Delta V = 0.64$~km,s$^{-1}$ are applied to the space velocity components.
	\footnote{LSR components are obtained via $(U_{LSR}, V_{LSR}, W_{LSR}) = (U + U_{\odot}, V + V_{\odot}, W + W_{\odot})$. The effects of Galactic differential rotation on $U$ and $V$ are estimated using $\Delta U = Ad \cos b \cos l \sin 2l - d \cos b \sin l (A \cos 2l + B)$ and $\Delta V = Ad \cos b \sin l \sin 2l + d \cos b \cos l (A \cos 2l + B)$. Note that $W$ velocities are assumed to be unaffected, under the assumption that the stars reside near the Galactic plane. Here, the Oort constants  $A=15.1\pm0.10$, $B=-13.4\pm0.10$ \citep{li2019}, $d$ is the distance of the cluster from the Sun, $l$ and $b$ are Galactic coordinates, respectively.}

	We adopt $R_{\odot}=8.2\pm0.1$ kpc \citep{bg16}, a circular LSR velocity
	$V_{LSR}=$239~km\,s$^{-1}$ \citep{bru11}, and $d=575\pm58$ pc as the distance of NGC 6793. The heliocentric Cartesian 
	positions $(x',y',z')$ and the LSR–corrected velocities
	$(U_{LSR},V_{LSR},W_{LSR})$ are then transformed into the 
	Galactic rest frame (GSR), yielding $(x,y,z)$ and $(V_x,V_y,V_z)$. The azimuthal velocity  ($V_{\Phi}$) (km s$^{-1}$) is estimated via
	
	\begin{equation}
		\begin{aligned} \label{eq:VPHI}
			V_{\Phi} =  \frac{x V_{y} - y V_{x}}{R}
		\end{aligned}
	\end{equation}
	
	The Galactic orbit of NGC~6793 was integrated using the \texttt{galpy} dynamical library \citep{Bovy2015} with the composite \texttt{MWPotential2014} gravitational potential. This potential includes a power-law bulge with an exponential cutoff, a Miyamoto--Nagai disk, and an NFW halo. Although the potential's internal parameters were originally calibrated assuming $R_{\odot} = 8.0$~kpc and $V_{LSR} = 220$~km\,s$^{-1}$ \citep[e.g.,][]{Dehnen1998, Bovy2013}, these values define only the scaling of the mass model and do not constrain the coordinate system used for orbital integration. Using the \texttt{galpy} code, the perigalactic and apogalactic distances ($R_{min}, R_{max}$), maximum vertical height ($z_{max}$), and orbital eccentricity ($ecc$) were obtained. The eccentricity is defined as:
	
	\begin{equation} 
		\begin{aligned}\label{eq-Vphi}
			ecc = \frac{R_{max}-R_{min}}{R_{max}+R_{min}}
		\end{aligned}
	\end{equation}
	
	The guiding or mean orbital radius ($R_m$) is calculated as $(R_{min} + R_{max})/2$. The orbits are  integrated using the kinematic parameters from Table~\ref{tbl-overall} for the cluster's estimated age (Table~\ref{tbl-astro-lit}). The orbital angular momentum component ($J_z$) has been derived from $J_z = xV_y - yV_x$. All parameters are summarized in Table~\ref{tbl-overall}. Uncertainties were estimated via a Monte Carlo approach by randomly sampling observational errors over 10,000 realizations.
	
	\begin{table*}[htb]
		\tabularfont
		\centering
		\caption{
			Kinematic, dynamical and spatial parameters of NGC~6793. All distances are in kpc, velocities in km s$^{-1}$, and $P$ in Myr.}
		\label{tbl-overall}
			\begin{tabular}{lcccc}
				\topline
				Parameter & $R_{\odot}=8.2$  &  $R_{\odot}=8.5$ & $R_{\odot}=8.2$ & $R_{\odot}=8.5$ \\
				& $V_{LSR}=239$ & $V_{LSR}=239$ & $V_{LSR}=220$ & $V_{LSR}=220$ \\
				\midline
				$U$          & -21.70 $\pm$0.63 &-22.29 $\pm$1.68 &-21.89 $\pm$1.49 & -21.04 $\pm$1.58 \\
				$V$           & -9.05 $\pm$1.90 &-8.05 $\pm$2.63 &-7.71 $\pm$2.32 & -8.81 $\pm$2.35\\ 
				$W$       & -5.60 $\pm$0.22 &-5.96 $\pm$0.62 &-5.77 $\pm$0.54 & -5.46 $\pm$0.46\\ 
				$U_{LSR}$   & -10.60 $\pm$0.63 & -11.19$\pm$1.68 & -10.79$\pm$1.49 & -9.94 $\pm$1.58 \\
				$V_{LSR}$ & 3.19 $\pm$1.90 & 4.19$\pm$2.63 & 4.53$\pm$2.32 & 3.43 $\pm$2.35\\
				$W_{LSR}$   & 1.65 $\pm$0.22 & 1.29$\pm$0.62 & 1.48$\pm$0.54 & 1.79 $\pm$0.46 \\
				$V_{\Phi}$    & -241.16 $\pm$ 1.83  & -242.03$\pm$2.56  & -223.40$\pm$2.23 & -222.54 $\pm$2.32\\ 
				$ecc$           & 0.08 $\pm$0.01 & 0.08$\pm$0.01 & 0.09$\pm$0.01 & 0.08 $\pm$0.01 \\
				$R_{min}$     & 7.37 $\pm$0.04 & 7.63$\pm$0.12& 7.35$\pm$0.10 & 7.70 $\pm$0.11\\
				$R_{max}$   & 8.62 $\pm$0.07 &9.00$\pm$0.12 & 8.73$\pm$0.13 & 8.96 $\pm$0.11\\
				$R_m$       & 7.99  $\pm$ 0.05 & 8.31$\pm$0.08 & 8.04$\pm$0.07 & 8.33 $\pm$0.08\\	
				$z_{max}$  & 0.06 $\pm$0.01 & 0.07 $\pm$0.01 & 0.07 $\pm$0.01 & 0.07 $\pm$0.01\\
				$R_{in}$    & 7.94 $\pm$0.45 &7.79$\pm$0.49 & 7.39$\pm$0.03 & 8.02 $\pm$0.50 \\
				$R_{GC}$   & 7.90 $\pm$0.02 & 8.18$\pm$0.04 & 7.88$\pm$0.03 & 8.21 $\pm$0.03 \\
				$J_z$         & 1905.92 $\pm$10.86 & 1979.53$\pm$18.20& 1761.35$\pm$14.34 & 1827.42 $\pm$16.95\\
				$P_{\phi}$          & 206.67 $\pm$ 1.75 &214.06$\pm$1.80 & 223.74$\pm$2.46 & 232.36 $\pm$2.52\\
				$N_{Rev,\phi}$     & 2.54  $\pm$ 0.02 & 2.34$\pm$0.04 & 2.24$\pm$0.05 & 2.15 $\pm$0.03 \\			
				$P_{rad}$       & 152.56 $\pm$ 1.40 &158.51$\pm$1.71 & 166.34$\pm$1.90 & 172.12 $\pm$1.71\\
				$N_{osc,rad}$     & 3.29  $\pm$ 0.05 & 3.17$\pm$0.03 & 3.01$\pm$0.03 & 2.91 $\pm$0.03 \\
				\hline
			\end{tabular}
	\end{table*}

	Using Equation~\ref{eq:VPHI}, we obtained the value of $V_{\Phi}\sim-241$ km s$^{-1}$. We note that $V_{\Phi}< 0$ means prograde. With the values of $V_{\Phi}$, eccentricity, $J_{z}$ and the other parameters in Table~\ref{tbl-astro-lit}, NGC~6793 reflects the characteristics of the Galactic disk. The x-y~(kpc) plane is known as projected on to the Galactic plane, whereas z-R (kpc) is the meridional plane.
	
\begin{figure*}[!t]
		\centering
		\includegraphics[width=0.8\linewidth]{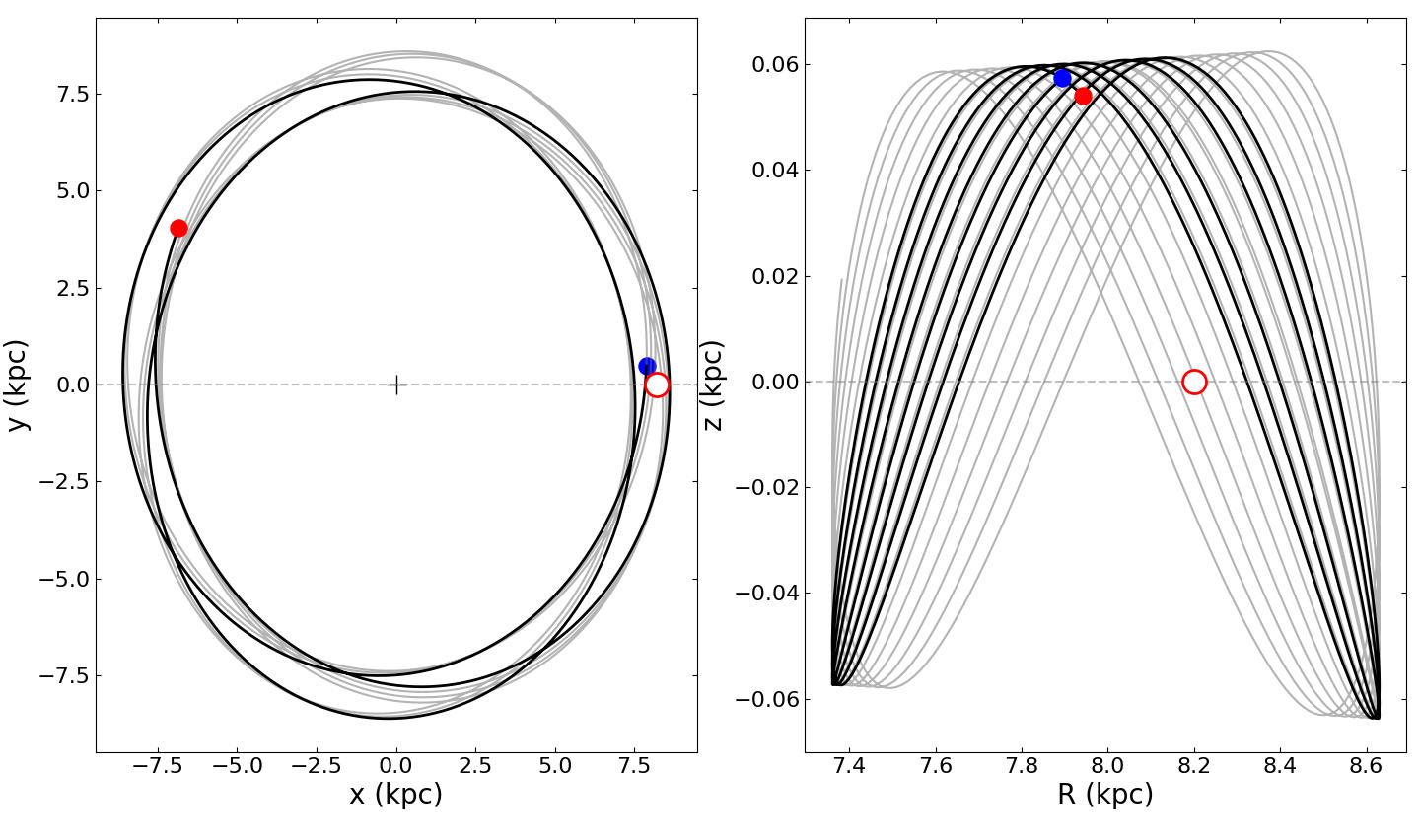}
		\caption{Galactic orbit of NGC~6793. The trajectories represent the paths traveled by the OC through its age. The filled red/blue dots show its initial/present day position. The open red circle shows the Sun's position.}
		\label{fig-orbit}
	\end{figure*}

	The dynamical history of NGC~6793 is characterized by two distinct orbital frequencies, as illustrated in Fig.~\ref{fig-orbit}. On the $x-y$ plane, the cluster follows an circular path ($ecc=0.08$) and an azimuthal period of $P_{\phi}\sim207$~Myr. Given its age 525 Myr, the cluster has completed $N_{Rev,\phi}\sim3$ revolution around Galaxy center.
	
	Simultaneously, the orbit in the $z-R$ plane exhibits characteristic boxy trajectories, with the cluster oscillating vertically within a confined meridional space defined by $7.37 < R \leq 8.62$~kpc and $|z| \leq 0.06$~kpc. The radial (epicyclic) period is significantly shorter than the azimuthal period, measured at $P_{rad} \sim153$~Myr. Over its evolution, the cluster has completed $N_{osc,rad}\sim 3$ full radial oscillations. Visually, this corresponds to approximately 6.88 individual radial crossings (excursions between the inner ($R_{min}$) and outer radial limits ($R_{max}$)), as seen by the overlapping loops in the $R-z$ projection. 
		Its initial ($t=-525$~Myr) and present day position in our Galaxy are displayed in Fig.~12(b) with the red and blue squares, respectively. NGC~6793's closest approaches to the sun is d$\sim0.58$~kpc.	
	
	To assess the robustness of our analysis, we compared our primary model ($R_{\odot}=8.2$~kpc, $V_{LSR}=239$~km~s$^{-1}$) against configurations utilizing $V_{LSR}=220$~km~s$^{-1}$ \citep{McMillan2017} and $R_{\odot}=8.5$~kpc \citep{Kerr1986}. The vertical extent ($z_{max} \approx 0.06-0.07$~kpc) and space velocity components ($U, V, W$) remained largely insensitive to these variations. While $V_{\Phi}$ and angular momentum ($J_{z}$) scale predictably with the choice of $V_{LSR}$, the orbital eccentricity ($ecc$) remains stable at $\sim 0.08$.

	\section{Cluster Mass and Mass Function Slope}
	
\begin{figure}[!t]
		\centering
		\includegraphics[width=0.85\linewidth]{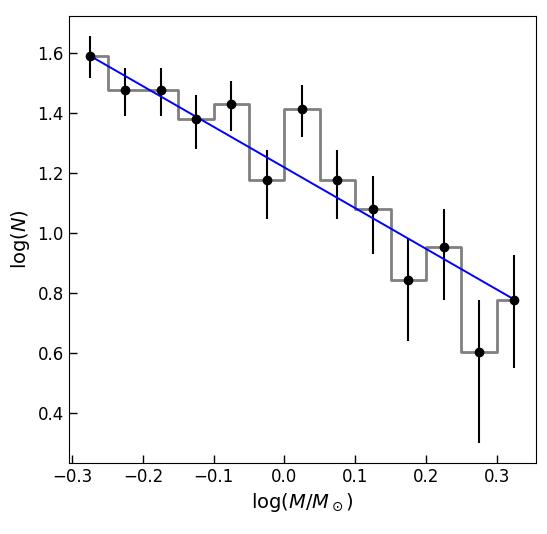}
		\caption{The mass function of NGC~6793. The vertical error bars as the Poisson noise are proportional to $\sqrt{N}$.}
		\label{fig-massfunc}
	\end{figure}

	The masses of NGC~6793 members have been determined using the PARSEC isochrone for $Z=0.024$ and $\log(Age/\text{yr})=8.72$. The $G$ magnitudes are converted into absolute magnitudes using the cluster’s reddening and distance modulus (Table~\ref{tbl-astro-lit}). The total mass has been  calculated for the 291 members identified with membership probability $P > 0.5$ as $242.13 \pm 5.33~M_\odot$. The mass function (MF) represents the distribution of stellar masses, with its slope ($\chi$) fitted to a power-law given by:
	
	\begin{equation}
		\begin{aligned}
			\frac{dN}{dM}=\phi(M)=M^{-(1+\chi)} \\[1ex]\label{eq:dNdM}
		\end{aligned}
	\end{equation}

	where $dN$ is the number of cluster members in the mass interval $dM$ centered at mass $M$. To determine the MF slope, we divided the stellar masses into logarithmic bins and tested several binning intervals (0.03, 0.05, 0.07, 0.09, and 0.10~dex). For each case, we derived the corresponding slope $\chi$ and the reduced chi-squared $\chi^2_{\nu}$ (Table~\ref{tab:bin_sensitivity}). Bin widths of 0.05 and 0.09~dex yielded $\chi^2_{\nu}$ values closest to unity (1.14 and 0.78, respectively), and both provided statistically consistent slopes ($1.50 \pm 0.17$ and $1.45 \pm 0.15$). Given that the 0.05~dex binning provides the best goodness-of-fit, we adopt its slope, $\chi = 1.50 \pm 0.17$, as our final result. The mass function derived from the full membership ($P > 0.5$) using this optimal binning is illustrated in Fig.~\ref{fig-massfunc}.
	
	We further examined the sensitivity of the MF slope ($\chi$) to the adopted mass interval by fitting several sub-ranges (Table~\ref{tab:range_sensitivity}). These tests involved excluding the lowest masses, the highest masses, and both extremes; when the highest masses are excluded, the MF slope becomes flatter ($\chi = 0.84 \pm 0.37$). For the exclusion of the lowest masses, the MF becomes steeper ($\chi = 2.27 \pm 0.47$). This variation is consistent with the expected effects of dynamical mass segregation for  an intermediate-age cluster. The intermediate interval yields a steeper slope than the full range; however, the full mass range ($-0.3 \leq \log(M/M_\odot) \leq 0.4$) provides the most statistically robust fit with $\chi^2_{\nu} = 1.14$ and aligns closely with the Kroupa IMF value ($\chi = 1.3 \pm 0.7$) for this regime \citep{Kroupa2001}. This stability throught the tested intervals confirms the statistical reliability of the adopted MF parameters for NGC~6793. Applying Equation~\ref{eq:dNdM} to the observations in Fig.~\ref{fig-massfunc} gives $\log(N/\Delta M)=(-2.50\pm0.17)\log(M/M_{\odot})+0.81$.
	
	\begin{table}[ht]
		\centering
		\caption{Sensitivity of the MF Slope ($\chi$) to Binning Choice.}
		\label{tab:bin_sensitivity}
			\begin{tabular}{ccc}
				\hline
				Bins & Slope ($\chi$) & $\chi^2_{\nu}$ \\
				\midrule
				0.03   & $1.77 \pm 0.23$ & 2.36 \\
				0.05  & $1.50 \pm 0.17$ & 1.14 \\
				0.07 & $1.45 \pm 0.19$ & 2.05 \\
				0.09 & $1.45 \pm 0.15$ & 0.78 \\
				0.10 & $1.50 \pm 0.18$ & 1.48 \\
				\hline
			\end{tabular}
		
	\end{table}
	
	\begin{table}[ht]
		\centering
		\caption{
				Sensitivity of the MF Slope ($\chi$) to Fitted Mass Interval.}
		\label{tab:range_sensitivity}

			\begin{tabular}{ccc}
				\toprule
				Mass Range ($\log(M/(M_\odot))$) & Slope ($\chi$) & $\chi^2_{\nu}$ \\
				\midrule
				-0.3 -- 0.4 (Full) & $1.50 \pm 0.17$ & 1.14 \\
				-0.1 -- 0.2 (Core) & $1.98 \pm 0.63$ & 1.68 \\
				0.0 -- 0.4 (No Low End) & $2.27 \pm 0.47$ & 0.64 \\
				-0.3 -- 0.1 (No High End) & $0.84 \pm 0.37$ & 1.16 \\
				\hline
			\end{tabular}
		
	\end{table}
	
	Additionally, we estimated the effect of unresolved binaries on the MF following \citep{Hunt2024}. We used a PARSEC isochrone with the cluster parameters ($\log(\mathrm{Age}) = 8.72$, $Z = 0.024$). First, we calculated the total mass assuming all stars are single. Then, we repeated the calculation including an unresolved binary fraction of 0.3. The total mass increased from 242.13~$M_\odot$ to about 310~$M_\odot$ ($\sim28\%$). This shows that unresolved binaries can significantly bias the mass. Therefore, the observed mass of 242.13~$M_\odot$ should be considered a lower limit.

	Variations in the adopted mass–luminosity relation (e.g., between PARSEC and MIST models, or due to different ages and metallicities) introduce small differences in the derived stellar masses, typically at the level of a few percent. The larger discrepancies mainly arise from the choice of the tidal radius, $R_{t}$, and the resulting number of selected members. In this study, we adopted $R_{t} = 48.93$ arcmin, while other works \citep[e.g.,][]{Tarr2022, hunt2023, alm2025} use larger values, leading to higher member counts and total masses. For example, the difference between our mass and that of A25 ($394,M_{\odot}$ for 1017 members) is consistent with this difference in membership selection. In addition, the pyUPMASK algorithm with GUMM cleaning provides robust outlier rejection but may be more conservative in including distant members in sparse clusters. Finally, our binary simulation shows that unresolved binaries increase the total mass by $\sim28\%$, indicating that our reported mass should be considered a lower limit. The final mass range, MF slope, total mass, mean mass, and number of members for NGC6793 are summarized in Table\ref{tbl-dynamic} for both $R_{\mathrm{lim}}$ and $R_{h}$ (see Section~\ref{sec-dynamics} for details on $R_{h}$).

\begin{figure}[!t]
		\centering
		\includegraphics[width=0.95\linewidth]{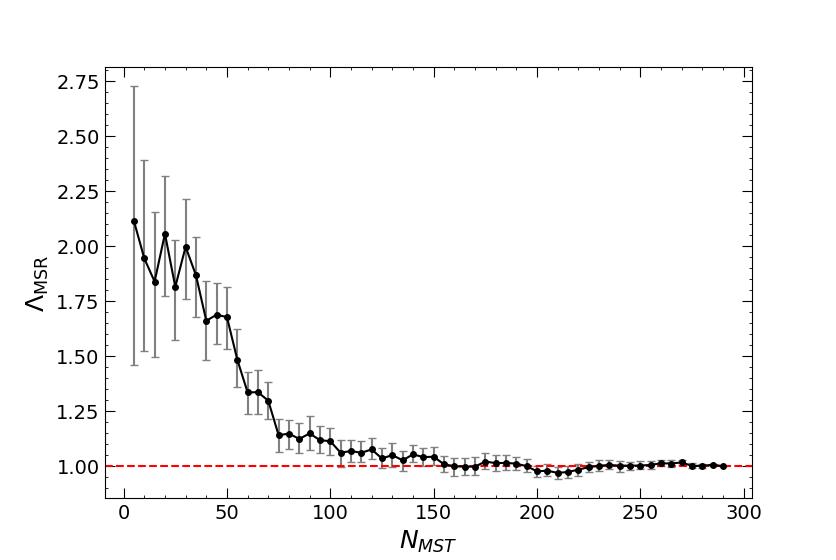}
		\caption{Mass segregation ratio $\Lambda_{\rm MSR}(N)$ for NGC~6793 as a function of the $N$ most massive stars. The horizontal red dotted line denotes the limit $\Lambda_{\rm MSR}=1$, which indicates no mass segregation. Gray bars represent CI.}
		\label{fig-mass-seg}
	\end{figure}

	Following the methodology of \cite{al2009a}, we estimated the mass segregation ratio, $\Lambda_{MSR}$, using Equation~\ref{eq:MSR} for the 291 identified cluster members. Comprehensive definitions of the terms in this equation are provided in \cite{al2009a} and \cite{akk24}. The resulting distribution of $\Lambda_{MSR}$ as a function of the $N$ most massive stars ($N_{MST}$) is illustrated in Fig.~\ref{fig-mass-seg}.

	\begin{align}\label{eq:MSR}
		\begin{aligned}
			\Lambda_{MSR} (N) = \frac{<l_{random}>}{l_{massive}} \pm \frac{\sigma_{random}}{l_{massive}},\\[1ex]
		\end{aligned}
	\end{align}
	
	We find clear evidence of mass segregation, with $\Lambda_{MSR}$ reaching a maximum of 2.15 for $N_{MST} \leq 100$. This result is supported by the 68\% confidence intervals (CIs), which were determined by bootstrapping the distribution 1000 times at each $N$. The upper and lower confidence limits correspond to the 84th and 16th percentiles of the bootstrap distribution, respectively, and are plotted as gray error bars in Fig.~\ref{fig-mass-seg}. Within these uncertainty limits, no further signatures of significant mass segregation or inverse mass segregation are observed beyond $N=100$. To further evaluate the statistical significance of this effect, we derived Monte Carlo $p$-values using the Anderson–Darling k-sample test ($P_{AD}$) for the $N_{MST} \leq 100$ regime. The $p$-values in this interval range between 0.001 and 0.04, confirming that the mass segregation detected by $\Lambda_{MSR}$ is statistically significant.

	\section{Dynamical parameters}\label{sec-dynamics}
	
	$t_{rlx}$ gives the time required for the stars in the core/halo to travel from one end of these regions to the other. The two $t_{rlx}$ equations are used in the literature, one for the overall cluster radius  ($R_{lim}$)  and the other for the central regions of clusters ($R_{h}$). Here, we adopt $R_{t}=R_{lim}$.
	The $t_{rlx}$ equation which considers $R_{lim}$  is used in the work of \cite{Bonatto2005}, and it is obtained from Equation~\ref{eq-trlx}.
	
	\begin{equation}\label{eq-trlx}
		\begin{aligned}
			t_{rlx_{1}}\approx0.04\left(\frac{N}{ln N} \right)\left(\frac{R_{lim}}{1pc}\right)
		\end{aligned}
	\end{equation}

	We adopted an internal velocity dispersion of $\sigma_{v}\approx1~{km\,s^{-1}}$, following the classical open-cluster estimate of \cite{Girard1989} and the value commonly used in OC dynamical studies such as \cite{Bonatto2011}. Observations show that intermediate-age and old open clusters typically have 	$\sigma_{v}\sim0.5$--$2~\mathrm{km\,s^{-1}}$ \citep[e.g.][]{Geller2010}, making our choice appropriate and significantly lower than the $\sigma_{v}\approx3~\mathrm{km\,s^{-1}}$ characteristic of hotter stellar systems \citep{Binney1998}. This dispersion value is also consistent with the findings of \cite{Belwal2025}, who likewise adopt $\sigma_{v}\approx1~\mathrm{km\,s^{-1}}$ as a representative value for intermediate-age open clusters. $N=291$ is the number of stars located inside $R_{t}<$8.34~pc (Table~\ref{tbl-struct}) and used in mass calculation. $t_{rlx}$ and $\tau$ have been obtained as  $t_{rlx_{1}}=$17.11$\pm$1.75~Myr and  $\tau_{1}=30.68\pm4.33$, respectively (Col.~2 of Table~\ref{tbl-dynamic}).
	
	\begin{table*}[htb]
		\tabularfont
		\centering
		\caption{The obtained parameters of mass information, dynamical indicators, and migration parameters for NGC~6793. The numbers in brackets in Cols.~2--3 belong to the values of Angelo et al. (2025).}
		\label{tbl-dynamic}
			\begin{tabular}{l c | l c | l c | l c}
				\topline
				\multicolumn{2}{c|}{Mass information} &
				\multicolumn{2}{c|}{Dynamical parameters} &
				\multicolumn{2}{c|}{Dynamical indicators} &
				\multicolumn{2}{c}{Migration parameters} \\
				\midline
				Mass range \,($M_{\odot}$) & $0.35$--$2.56$ & $R_h$\,(pc) & $3.20\pm0.01$~(1.70) & $c$ & $0.71$~(0.87) & $R_{guide}$\,(kpc) & $7.90 \pm0.08$ \\
				MF slope \, ($\chi$) & $1.50 \pm 0.17$ & $R_J$\,(pc) & $8.44\pm0.08$~(6.20) & $R_{c}/R_{h}$ & $0.51$~(0.53) & $R_{birth}$\,(kpc) & $4.28 \pm0.29$ \\
				& & & & $R_{h}/R_{t}$ & $0.38$~(0.26) & $d_{mig}$\,(kpc) & $3.62\pm0.29$ \\
				& & & & $R_{h}/R_{J}$ & $0.38$~(0.27) & & \\
				& & & & $R_{t}/R_{J}$ & $0.99$~(1.06) & & \\
				& & & & $R_{J}/R_{c}$ & $5.21$~(6.89) & & \\
				\hline
				& & \multicolumn{2}{c|}{For $R_{t}$} & & & & \\
				\midline
				cluster mass \,($M_{\odot}$) & $242.13 \pm 5.33$ & $t_{rlx1}$\,(Myr) & $17.11\pm1.75$ & & & & \\
				Members \, (N) & $291.0 \pm 17$ & $\tau_{1}$ & $30.68 \pm 4.33$ & & & & \\
				& & $\log(\tau_{1})$ & $1.49\pm0.06$ & & & & \\
				& & $t_{diss}$ & $690.40 \pm 42.78$~Myr & & & & \\
				\hline
				& & \multicolumn{2}{c|}{For $R_{h}$} & & & & \\
				\midline
				cluster mass \,($M_{\odot}$) & $162.26 \pm 2.38$ & $t_{rlx2}$\,(Myr) & $38.79 \pm 1.70$~(38) & & & & \\
				Mean Mass \,($M_{\odot}$) & $0.90 \pm 0.07$ & $\tau_{2}$ & $13.53 \pm 1.44$~(11.77) & & & & \\
				Members \, (N) & $180 \pm 13$ & $\log(\tau_{2})$ & $1.13\pm0.05$ & & & & \\
				\hline
			\end{tabular}
		\tablenotes{
			All masses are in $M_{\odot}$, distances in pc/kpc, and times in Myr. 
			Values in brackets are from Angelo et al. (2025). 
			$R_{c}$ and $R_{t}$ are the core and tidal radii derived in Sect.~\ref{sec:structural_membership}; 
			$R_{h}$ is the half-mass radius (Fig.~\ref{fig-hlf}). 
			The Jacobi radius $R_{J}$ is computed using Eqs.~(\ref{eq-RJ})–(\ref{eq-freq}). 
			The relaxation times $t_{rlx1}$ and $t_{rlx2}$ follow Eqs.~(\ref{eq-trlx}) and (\ref{eq-trlx2}). 
			$\tau = \mathrm{Age}/t_{rlx}$, $c=\log(R_{t}/R_{c})$, 
			and the guiding radius is $R_{guide} = J_{z}/V_{\Phi}$, with $d_{mig} = R_{guide} - R_{birth}$.
		}
	\end{table*}
	
	As an indicator of the dynamical evolution, the evolutionary parameter is estimated from the relation $\tau=Age/t_{rlx}$ by using its age (525~Myr) (Table~\ref{tbl-astro-lit}).  Mass segregation (known as migrating from the cores to outer parts) is directly related to $t_{rlx}$ and  $\tau$ \footnote{$\tau$ also depends on both age and $t_{rlx}$. There is a negative relationship between $t_{rlx}$ and $\tau$. Large $\tau$ and small $t_{rlx}$ mean advanced mass segregation, accordingly, small $\tau$ and large $t_{rlx}$ mean small scale mass segregation. However, these relationships are not so direct. Dynamically young OCs show significant mass segregation, while dynamically older OCs do not. Mass segregation is not exclusively primordial and not only produced by two-body relaxation. As emphasized by A25 (see their sect.~5.5), the presence of mass segregation at young ages can significantly influence the advanced dynamical evolution of a cluster.}. We tested how different $\sigma_{v}$ would affect $t_{rlx}$, $\tau$, and $t_{diss}$. For $\sigma_{v}\approx2~{km\,s^{-1}}$, we found $t_{rlx}$=8.56$\pm$0.87 Myr and $\tau=61.36\pm8.65$, while for $\sigma_{v}\approx3~{km\,s^{-1}}$ we found $t_{rlx}$=5.70$\pm$0.58 Myr and $\tau=92.04\pm12.97$. Those results show that the values of our dynamical timescales are sensitive to $\sigma_v$. Our sensitivity analysis showed that while the exact value of the relaxation time depends on the assumed velocity dispersion, our main finding holds: the cluster is dynamically old enough for mass segregation to have occurred.
	
	\begin{figure}[!t]
		\centering
		\includegraphics[width=0.85\linewidth]{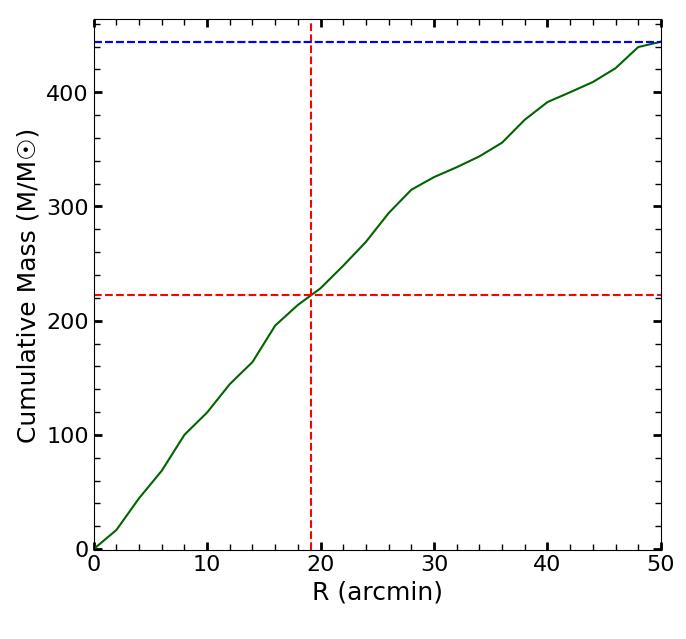}
		\caption{The cumulative mass versus radius of NGC~6793. The vertical line represents the half-mass radius corresponding to the half-mass (horizontal red dotted line). The upper blue dotted line denotes the total mass.}
		\label{fig-hlf}
	\end{figure}
	
	The half-mass relaxation or half-light relaxation time $t_{rlx}$ for central parts of the clusters is obtained from the relation by \cite{Spitzer1971}
	\begin{equation}\label{eq-trlx2}
		\begin{aligned}
			t_{rlx_{2}}=\frac{8.9\times10^5\sqrt{N}\times{R_{h}}^{3/2}}{log(0.4N)\times \sqrt{m}}\\[1ex]
		\end{aligned}
	\end{equation}
	
	where m and N refer to the mean mass and the number of the cluster members, respectively, taken  from Col.~1 of Table~\ref{tbl-dynamic}. $R_{h}$ is defined as the radius from the core that contains half the total mass of the cluster.  From the cumulative mass in dependence of the radius (Fig~\ref{fig-hlf}), $R_{h}$ of NGC~6793 is obtained as 19$^{\prime}$.14$\pm$0$^{\prime}$.03, which corresponds to  3.20$\pm$0.01~pc (Col.~2 of Table~\ref{tbl-dynamic}). Thus, we give the number of the members and mean mass within $R_{h}<19^{\prime}.14$ as $180$ and $0.90 M_{\odot}$ (Col.~1 of Table~\ref{tbl-dynamic}). From Equation~\ref{eq-trlx2}, we obtain $t_{rlx_{2}}=$38.79$\pm$1.70 Myr, and thus $\tau_{2}=$13.53$\pm$1.44.
	
	The initial mass ($M_{ini}$), ambient density ($\rho_{amb}$), and dissolution time ($t_{diss}$) for NGC~6793 were calculated following Equations~13--15 of \cite{Angelo2023} and Equation~10 of A25. For these calculations, we adopted $Z=0.024$ (Table~\ref{tab:litcompbig}), a cluster mass of $M_{cl}=242.13~M_{\odot}$ (Table~\ref{tbl-dynamic}), an eccentricity of $ecc=0.08\pm0.01$ and a Galactocentric distance of $R_{GC}=7.90\pm0.02$~kpc (Table~\ref{tbl-overall}). The constant "810" in the dissolution equation refers to the timescale required for the cluster to lose $95\%$ of its initial mass \citep{Lamers2005}. Consequently, we obtained $M_{ini} = 1462 \pm 152~M_{\odot}$, $\rho_{amb} = 0.13 \pm 0.01~M_{\odot}~\mathrm{pc}^{-3}$, and $t_{diss} = 690.40 \pm 42.78$~Myr. In comparison, A25 reported similar values for background density ($\rho_{amb}=0.14~M_{\odot}~\mathrm{pc}^{-3}$) and dissolution time ($t_{diss} \approx 700$~Myr). Their initial mass ($M_{ini}=1615~M_{\odot}$) is slightly higher than ours, while their current cluster mass is $\approx 63\%$ larger ($M_{cl}=394~M_{\odot}$). These discrepancies in initial and current masses arise primarily because A25 identified a significantly larger number of cluster members ($N=1017$) from Gaia DR3 data and adopted a higher eccentricity ($e=0.127$) from \cite{Tarr2021}. These differences in membership and orbital parameters naturally lead to the larger mass and longer dissolution time-scales found in their study.
	
	Its Jacobi tidal radius $R_{j}$ is estimated as 8.44$\pm$ 0.08 pc, using the following equation given by \cite{por2010},

	\begin{equation}\label{eq-RJ}
		R_{j} = \left( \frac{G M_{cl}}{2 \omega^{2}} \right)^{1/3}
	\end{equation}

	where $\omega$ is the orbital angular velocity, and is calculated by
	\begin{equation}\label{eq-freq}
		\omega = \frac{V_{\phi}}{R_{GC}}
	\end{equation}

	The concentration parameter is estimated from the relation $c=\log (R_{t}/R_{c})$. According to \cite {pia17a,pia17b},  $c$ is the efficiency of heating on cluster stars caused by dynamic interactions in the center of the clusters. This parameter is almost negatively correlated with $R_{h}/R_{j}$ (or $R_{h}/R_{t}$) and increases with age as a result of the dynamic evolution of the OCs \citep{Angelo2020,Angelo2021}.
	The birth radius ($R_{birth}$) of NGC~6793 is estimated  with the help of the current metallicity gradient based on young OCs, the model by \cite{min18} for the time evolution of the Galactic  Interstellar Medium (ISM) metallicity gradient, its metal abundance and age (Table~\ref{tbl-astro-lit}). Its guiding radius and migration distance are obtained from the relations $R_{guide}=J_{z}/V_{\Phi}$ and $d_{mig} = R_{guide}-R_{birth}$, respectively. All the obtained dynamical parameters are summarized in Table~\ref{tbl-dynamic}.

	
	\section{Discussion and Conclusion}\label{sec-discussion}
	Within the error limits our $E(B-V)=0.24\pm0.02$~mag is in concordance with the values (Table~\ref{tbl-astro-lit}) obtained from the CMDs at a level of $\Delta E(B-V)= 0.02$ to $0.09$, except for 0.45 value of \cite{He2022}. Moreover, the reddening value of this study is based on $U$-filter observation. Our distance value from bayesian method and simultaneous multi-CMD isochrone fitting is $575\pm58$~pc. This distance is in good agreement with the value  derived from Gaia DR3 trigonometric parallax (586$\pm$10~pc) (Table~\ref{tbl-propermotion}). 
	
	With the age $525\pm51$~Myr ($\log(Age/yr)=8.72\pm0.04$), NGC~6793 is an intermediate aged OC. \cite{sung13} give the age range of the intermediate-age OCs as 10--700~Myr ($\log(Age/yr)=$ 8.00--8.85).  Within the error limits our age is in concordance with the values of A25, \cite{alm2023}, \cite{dia21}, \cite{cantat2020}, and \cite{Tasdemir2025}. Note that some authors give a relatively young age, 71-309 Myr as compared to our detect.  The small/large discrepancies in ages and distance moduli/distances (Table~\ref{tbl-astro-lit}) stem from the adopted isochrones in terms of $Z$ heavy element abundances (Col.~4 of Table~\ref{tbl-astro-lit}) and the reddening values from the CMDs or CCs, as well as different membership techniques, as emphasized by \cite{kar2023}. Note that all authors in Table~\ref{tbl-astro-lit} use the same PARSEC isochrones.

	\begin{figure}[!t]
		\centering
		\includegraphics[width=1.02\linewidth]{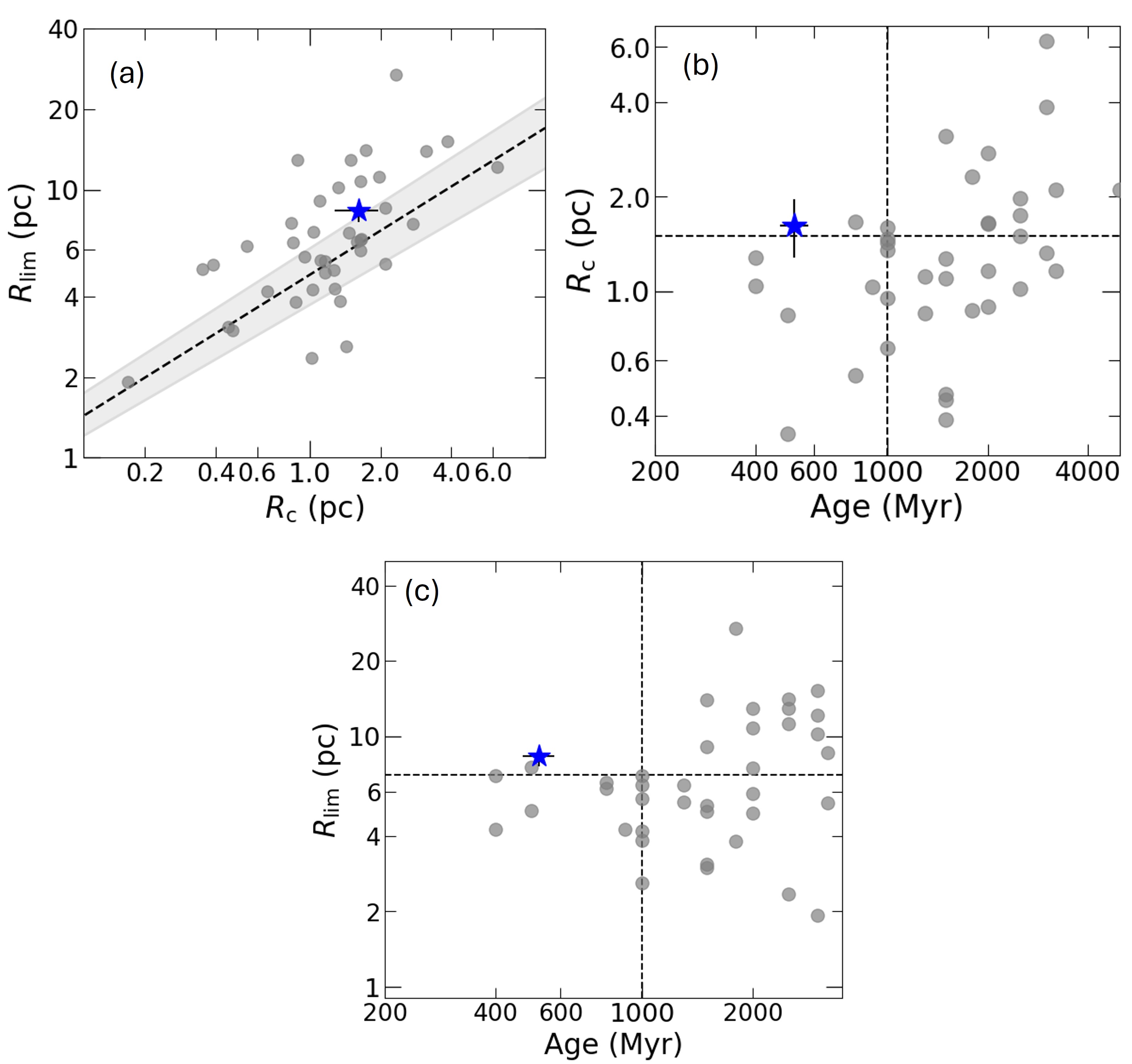}
		\caption{The location of NGC~6793 (blue star) in $R_{\rm lim}$ versus $R_c$ (panel~a) and $(R_c,~R_{\rm lim})$ versus Age (panels~b--c). The relation (solid line) and its $1\sigma$ uncertainty as the shaded gray area in panel~(a). The horizontal and vertical dotted lines show the separation limits of the small/large sized OCs and the bifurcation at $\approx1$ Gyr, respectively. The grey dots denote the OCs of Gunes et al. (2017). $R_{2}$ and $R_{4}$ regions represent the expanding and shrinking regions of the OCs.}
		\label{fig-dynamics1}
	\end{figure}

	NGC~6793 resides in the first quadrant ($\ell=56^{\circ}$) and $R_{GC}=$~$7.90$~kpc (inside the solar circle) together with its relatively small mass ($\sim242~M_{\odot}$). It has a close position to the Galactic disc (vertical distance $z\sim$60~pc). In conjunction with this, during its three rotations (Table~\ref{tbl-overall}) around the Galactic center, it feels external tidal perturbations such as Galactic potential, Giant Molecular Clouds (GMCs), tidal effects with the spiral arm and the Galactic disk, which are quite efficient in the Galactic center directions towards its dissolution.

	Its steep $\chi =1.50$ means that its low mass stars outnumber its massive ones. Its age (525~Myr) is larger than its relaxation time (t$_{rlx_{1}}=$17.11~Myr) (Table~\ref{tbl-dynamic}). Therefore, it is dynamically relaxed. When the evidences of its cluster dimensions, steep MFs and large $\tau_{1}=30.68$ are put together,  NGC~6793 appears to have lost its low mass star content from its outer skirts owing to the significant external tidal effects. In this sense, NGC~6793 presents a sign of mild mass segregation. This is further supported by the mass segregation ratio (Fig.~\ref{fig-mass-seg}), which shows clear signatures of segregation for $N \leq 100$, reaching a maximum value of $\Lambda_{MSR}^{max} = 2.11$ at $N = 5$.

\begin{figure}[!t]
		\centering
		\includegraphics[width=1.02\linewidth]{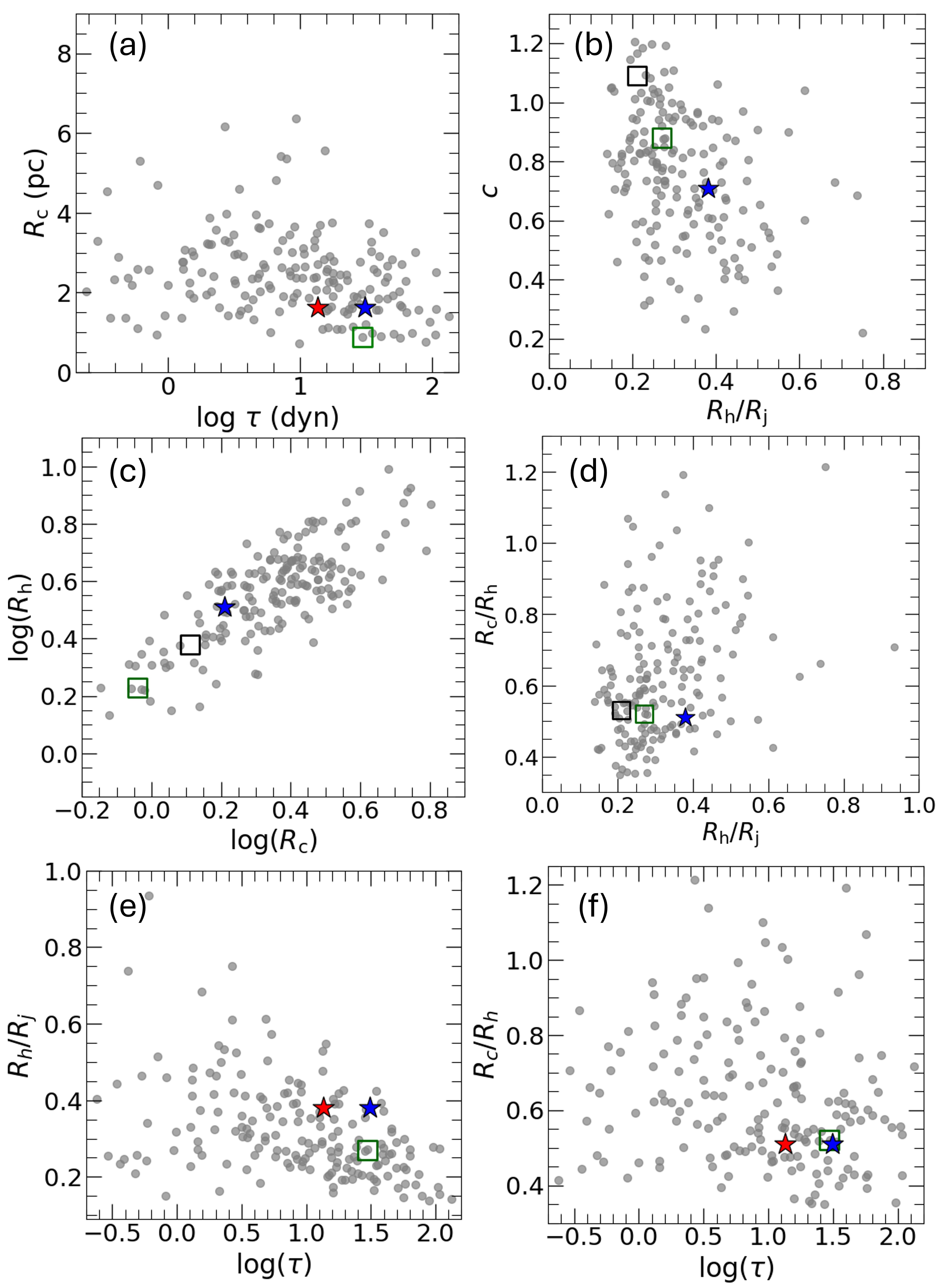}
		\caption{The position of NGC~6793 for $\log(\tau_{1})$ (blue star symbol) and $\log(\tau_{2})$ (red star symbol)) in the relations of $R_c$ versus $\log(\tau)$ (panel~a), $c$ versus $R_h/R_j$ (panel~b), $\log(R_h)$ versus $\log(R_c)$ (panel~c), $R_c/R_h$ versus $R_h/R_j$ (panel~d), and $R_h/R_j$, $R_c/R_h$ versus $\log(\tau)$ (panels~e--f). The gray dots in the panels are from A25. Green and black square symbols denote the values of A25 and H23, respectively.}
		\label{fig-dynamics2}
	\end{figure}

	The dynamical evolution of NGC~6793 has been interpreted with the help of the dynamical indicators of \cite{Gunes2017} and A25, as is displayed in Figs.~\ref{fig-dynamics1}-\ref{fig-dynamics2}. For the meaning of the symbols, see the figure captions. As simplicity, the values of  $2000~M_{\odot}$ and $(R_{c},~R_{lim})=(1.5,~7)$ pc given by \cite{Gunes2017} are considered to classify the OCs as massive/less massive and relatively small/large sized. To ensure it is compatible with the \cite{Gunes2017}' sample OC (gray dots) in Fig.16, as a simple approach, we adopt $R_{t}=R_{lim}$. With the relatively moderate dimensions, $(R_{c}, R_{lim})=(1.62,~8.34)$~pc, NGC~6793 follows the increasing trend between $R_{c}$ and $R_{lim}$ (panel (a)), given by \cite{camargo2009,camargo2010, Gunes2017}. 
	Another possibility for NGC~6793 moderate sizes may be linked to its initial formation from its parent molecular cloud \citep{van91, camargo2009}. Initial formation conditions play a role in the dynamical evolution of the OCs.

	The position of NGC 6793 in the $R_{1}$ region (Fig.~\ref{fig-dynamics1} b–c) does not indicate signs of expansion and shrinkage.  Its locus ($R_{1}$) remains to the left of the bifurcation age, $\approx1$~Gyr\footnote{From the Galactic OC samples, \cite{camargo2009, camargo2010} and \cite{Gunes2017} give the bifurcation age as $\approx 1$~Gyr. The relationship between the cluster dimensions and age is the indicator of the survival and dissociation rates of the OCs \citep{camargo2009}}. According to A25, the loosely bound OCs with the relatively small $t_{diss}/t_{rlx}<7$ can be easily disrupted. With the values t$_{diss}$/t$_{rlx_{1}}$$=$40 and $\log R_{j}/R_{c}=$0.72 (Table~\ref{tbl-dynamic}), its locus is in reasonably concordance with the relatively compact group OCs (orange square symbols with R$_{GC}<$ 8~kpc) in Fig.14(a) of A25. In line with this, NGC~6793 with $R_{h}/R_{j}=0.38$ and $\log\rho_{amb}=$--0.89 obeys the positive correlation in Fig.~10(a) of A25, and these values place it within the relatively compact clusters in $R_{GC} < 8$~kpc. These all necessarily mean that its compact internal structure is stable against tidal disruption which is responsible for the combined effect of the tidal heating and two-body encounters.  When considered its t$_{diss}\sim$~690~Myr (Table~\ref{tbl-dynamic}) and its age,  NGC~6793 has survived for nearly 76$\%$ of its total lifetime. It lost 83$\%$ of initial mass (1462~$M_{\odot}$). In this respect, it could dissolve and disperse before it reaches $R_{4}$ region.

	Within the error limits, NGC~6793 almost follows the general trends in the panels of Fig.~\ref{fig-dynamics2}. Note that A25'sample includes the OCs with different evaporation regimes which do not follow the general trends. H23 also give the ($R_{c}$,~ $R_{h}$,~ $R_{t}$,~$R_{j}$)=(1.28, 2.70, 15.67, 11.45)~pc for NGC~6793. Here we consider their R50J as half-mass radius-R$_{h}$, and thus we obtained the dynamical ratios as $R_{c}/R_{h}=0.47$, $R_{h}/R_{t}=0.17$, $R_{h}/R_{j}=0.24$, and $c=$1.09, which are almost concordance with our detects (Cols.~2-3 of Table~\ref{tbl-dynamic}). Their detection also meets $R_{h}>R_{c}$ requirement. In panels~b--d their dynamical indicators (blue square symbol) almost obey the general trends, and thus support our conclusions.

	Here it is useful to specify the boundary values \footnote{According to \cite{fuk1995}, the OCs with $R_{h}/R_{j}>0.70$ may be quickly destroyed as a result of their instability and strong tidal effects. The OCs  with $R_{c}/R_{h}< 0.20$ are core-collapsed systems. Both $R_{h}/R_{j}$ and  $R_{h}/R_{t}$ are the indicators of tidal filling. As emphasized by A25,  $R_{t}$ is a more uncertain parameter in the King fits to the observations. Accordingly, it subjects to fluctuations of the stellar density. The tidal filling ratio $R_{h}/R_{j}$ is a ratio to evaluate the cluster dynamical state, since it is determined by both the internal evolution and the external tidal field. $R_{h}/R_{j}$ tends to decrease with $\log(\tau)$ as a consequence of the internal relaxation process, which causes the clusters’ main body to be progressively more compact.} for commenting the dynamical evolution of NGC~6793. 
	
	The $R_{c}$ and $\log\tau_{1}$ of NGC~6793 (Fig.17(a)) confirm the decreasing trend of A25 (see their Fig.~9). Following the arguments of A25 and \cite{heg2003}, we state that NGC~6793' core radius shrinks along its advanced dynamical evolution ($\log\tau_{1}=1.49$) (Table~\ref{tbl-dynamic}), as a consequence of mass segregation and evaporation of less massive stars in its core region. As stated by A25, $R_{c}$ values show smaller dependence with the external conditions compared to $R_{h}$ and $R_{t}$, implying that the clusters’ central structure is more sensible to the internal evolution.  This explains well its place  in Fig.~\ref{fig-dynamics1}(b).
	
	NGC~6793 with c$=$0.71 is a moderately concentrated OC with highly relaxed internal structure. There is almost a negative correlation between $c$ and $R_{h}/R_{j}$ with some scatter (panel~b). Consequently, it tended to relatively small $R_{h}/R_{j}=0.38$ because of the evaporation effects (see A25's Fig.12a-b). 
	It meets the requirement of compliance with the general trend with $R_{h}=3.20~pc > R_{c}=1.60~pc$   (panel~c). Our $R_{c}/R_{h}=0.51$ and $R_{h}/R_{j}=0.38$ ratios reconciliates better with the sample OCs of A25 (panel~d). However, our $R_{h}/R_{j}=0.38$ value is slightly larger than the ones of A25 and H23 (panel~d). For $R_{h}$ and $R_{j}$ (Table 11), A25 used a different method (see their work for the details).
	
	In panel~(d) the dynamical evolution of NGC~6793 is ruled by both internal two-body relaxation and external tidal perturbations. Its $R_{c}/R_{h}=0.51$ is responsible for two-body relaxation, mass segregation and core-collapse
	in its core region ($\log\tau_{2}=1.13$)~(panel~f). However, NGC~6793's $R_{h}/R_{j}=0.38$ implies that it is tidally influenced since it quite close to upper limit, in the range $0.12 < R_{h}/R_{j} < 0.35$, according to Fig.~2a of \cite{heg2003}. Probably, its $R_{h}/R_{j}$ was relatively large in the past, following the general trend of $R_{h}/R_{j}$ decreasing with $\log(\tau)$. In parallel with this, NGC~6793 allowed its internal mass distribution to relax across Jacobi radius $R_{j}=8.44$~pc (the point of being tidally filling) without being tidally disrupted ($R_{h}/R_{j}<0.70$). In this context, its current $R_{h}/R_{j}$ indicates that it exposed to external galactic tidal processes, thus lost its low-mass star content to the field, as is evident from its large ($\log\tau_{1}=1.49$) (panel~e). NGC~6793 is well located in the locus of highly-dynamically evolved clusters (panel~e), thus internal two-body relaxation contracts its main body as it evolves dynamically.  Regarding its interior dynamical evolution, the relaxation also forced the high mass stars to its central region, and vice versa its low mass star content to the periphery. Based on the outcomes of \cite{heg2003}, $R_{c}/R_{h}$ ratio is also expected to decrease as consequence of the dynamical effects.

        The OCs with $R_{t}/R_{j}\gtrsim~1$ are tidally overfilled. Whereas the OCs with $R_{t}/R_{J}<1$ keep their stellar content within its Jacobi radius \citep{Angelo2023}. In light of this information,  
		NGC~6793 with $R_{t}/R_{j} = 0.99$ appears to be in a tidally filling state, positioned precisely at its gravitational boundary. This indicates that while the cluster is not yet overfilled, it is on the verge of significant mass loss due to external tidal effects. At this limit, the stars approaching $R_{t}$ can be interpreted as an unbound "cluster corona"---escaping members that have not yet fully dispersed into the Galactic field, potentially forming the inner tidal tails \citep{Kupper2010}. In such cases, the \cite{King1962} model, which assumes a sharp tidal truncation, is forced to account for this extended population, leading to an $R_{t}$ value that reflects the observational boundary near the true gravitational limit.
	
	We obtain that NGC~6793 migrated about 3.62 kpc from its birth place (4.28 kpc). It has guiding radius (7.90 kpc) fairly near its current radius. Its relatively large migration rate is also compatible with the cluster’s moderate evaporation due to low-mass loss and advanced dynamical evolution (relatively high $\tau$).

	
	\vspace{-1em}
	
	\section*{Acknowledgements}
	
	We sincerely thank the editor and the anonymous reviewers for their careful reading of the manuscript and for their constructive and insightful comments, which significantly strengthened the quality and clarity of this work. We are grateful to Dr.~M.~Angelo for his valuable comments and scientific discussions regarding the dynamical evolution parameters of the cluster. Z.A. acknowledges the financial support provided by the \.{I}lim Yayma Vakf{\i} (\.{I}YV) scholarship. The open cluster data is based upon observations carried out at the Observatorio Astron\'{o}mico Nacional on the Sierra San Pedro M\'{a}rtir (OAN-SPM), Baja California, M\'{e}xico. This paper has made use of results from the European Space Agency (ESA) space mission \textit{Gaia} (\url{http://www.cosmos.esa.int/gaia}), with data processed by the \textit{Gaia} Data Processing and Analysis Consortium (DPAC). Funding for the DPAC has been provided by national institutions, in particular the institutions participating in the \textit{Gaia} Multilateral Agreement. This research has also made use of the WEBDA database, operated at the Department of Theoretical Physics and Astrophysics of the Masaryk University, Brno, and the SIMBAD and VizieR databases (\url{http://vizier.cds.unistra.fr/viz-bin/VizieR-2}).
	
	

	{}

\end{document}